\documentclass[fleqn,usenatbib,useAMS]{mnras}

\usepackage{graphicx}	
\usepackage{amsmath}	
\usepackage{amssymb}	
\usepackage{multicol}        
\usepackage{bm}		
\usepackage{pdflscape}	
\usepackage{placeins}
\usepackage{url}

\usepackage{journals}
\newcommand{\bs}[1]{\overline{#1}}
\def\bm#1{\ensuremath{\mathchoice{\mbox{\boldmath$\displaystyle#1$}}
{\mbox{\boldmath$\textstyle#1$}}
{\mbox{\boldmath$\scriptstyle#1$}}
{\mbox{\boldmath$\scriptscriptstyle#1$}}}}

\newcommand{\figref}[1]{Fig.~\ref{#1}}

\newcommand{\Figref}[1]{Fig.~\ref{#1}}

\newcommand{\be}{\begin{equation}}
\newcommand{\ee}{\end{equation}}
\newcommand{\eep}{\, .\end{equation}}
\newcommand{\eec}{\, ,\end{equation}}
\newcommand{\bea}{\begin{eqnarray}}
\newcommand{\bel}[1]{\be\label{#1}}

\newcommand{\eea}{\end{eqnarray}}

\newcommand{\back}[1]{\overline{#1}}

\newcommand{\rhoU}{\rho^U}

\newcommand{\gb}{\back{g}}

\renewcommand{\Gamma}{\varGamma}
\renewcommand{\Psi}{\varPsi}

\usepackage[T1]{fontenc}
\usepackage{ae,aecompl}

\usepackage{newtxtext,newtxmath}

\title[Zonal Winds and Gravity]{Linking Zonal Winds and Gravity II: explaining the equatorially antisymmetric gravity moments of Jupiter}

\author[W. Dietrich et al.]{Wieland Dietrich$^{1}$\thanks{Contact e-mail: {dietrichw@mps.mpg.de}},
Paula Wulff$^{1,2}$,
Johannes Wicht$^{1}$,
and Ulrich R.~Christensen$^{1}$
\\
$^{1}$Max Planck Institute for Solar System Research, Justus-von-Liebig-Weg 3, 37077, G\"ottingen, Germany\\
$^{2}$Georg August University, Institute for Geophysics, Friedrich-Hund-Platz 1, 37077 G\"ottingen}

\date{Last updated \today; in original form \today}

\pubyear{202-}

\begin{document}
\label{firstpage}
\pagerange{\pageref{firstpage}--\pageref{lastpage}}
\maketitle

\begin{abstract}
The recent gravity field measurements of Jupiter (Juno) and Saturn (Cassini) confirm the existence of deep zonal flows reaching to a depth of 5\% and 15\% of the respective radius. Relating the zonal wind induced density perturbations to the gravity moments has become a major tool to characterise the interior dynamics of gas giants. 
Previous studies differ with respect to the assumptions made on how the wind velocity relates to density anomalies, on the functional form of its decay with depth, and on the continuity of antisymmetric winds across the equatorial plane. Most of the suggested vertical structures exhibit a rather smooth radial decay of the zonal wind, which seems at odds with the observed secular variation of the magnetic field and the prevailing geostrophy of the zonal winds. Moreover, the results relied on an artificial equatorial regularisation or ignored the equatorial discontinuity altogether.
We favour an alternative structure, where the equatorially antisymmetric zonal wind in an equatorial latitude belt between $\pm 21^\circ$ remains so shallow that it does not contribute to the gravity signal. The winds at higher latitudes suffice to convincingly explain the measured gravity moments. Our results indicate that the winds are geostrophic, i.e. constant along cylinders, in the outer $3000\,$ km and decay rapidly below. The preferred wind structure is 50\% deeper than previously thought, agrees with the measured gravity moment, is compliant with the magnetic constraints and the requirement of an adiabatic atmosphere and unbiased by the treatment of the equatorial discontinuity.   
\end{abstract}




\section{Introduction}
The spacecrafts {\it Juno} and {\it Cassini} delivered high-accuracy measurements of the gravity potentials for Jupiter and Saturn \citep{Kaspi2018, Iess2018,Guillot2018,Galanti2019,Iess2019}, which provide valuable constraints on the interior structure and dynamics of their atmospheres. For the first time, it has been possible to resolve the tiny undulations in the gravity potential induced by zonal winds. This has ended the long-standing debate on whether the zonal winds on Jupiter and Saturn are shallow weather phenomena or reach deeper into the planets' convective envelopes. The gravity data suggest that the winds extend down to $5\%$ and $15\%$ of Jupiter's and Saturn's radii, respectively \citep{Kaspi2018,Galanti2019, Kaspi2020}. However, modelling the deep-reaching zonal mass fluxes on a gas planet with all their complexity and relating them to anomalies in the gravity field is a difficult, non-unique problem.

The recent attempts of constraining the deep-reaching winds with gravity data are based on simple parametrizations of the vertical wind structure. 
The observed surface winds are thus continued downward along cylinders that are aligned with the rotation axis. The resulting geostrophic flow is then multiplied with a parametrized radial profile to model the decay with depth. The parameters of this profile are constrained by the gravity measurements \citep{Kaspi2010,Kaspi2013,Kaspi2018,Kong2018,Galanti2019}. More recently,  
constraints from Jupiter's magnetic field \citep{Duer2019} or it's temporal evolution \citep{Galanti2020} have also been used in addition to this. 
However, some studies found it necessary to partially modify the zonal flow compared to what is observed at cloud level \citep{Kong2018, Galanti2020}  in order to match the modelled gravity anomalies to the observed ones.

Gravity moments of odd degree exclusively contain the impact of equatorially antisymmetric zonal flows, since they are not obscured by the rotational deformation of the planet as opposed to their even counterparts. The geostrophic, downward continuation of the antisymmetric zonal flows from each hemisphere yields opposite signs when they each reach the equatorial plane and hence introduces a problematic discontinuity or equatorial `step`. This has been reported to potentially bias the results \citep{Kong2016}. To circumvent this issue, several published studies, such as \citet{Kaspi2018, Galanti2020} calculate the antisymmetric gravity signal in each hemisphere, but ignore the resulting equatorial shear region. Alternatively \citet{Kong2017,Kong2018} suggested to smooth the equatorial region artificially.

Zonal winds induce perturbations in the gravity potential via pressure perturbations that change the density distribution. The governing leading order Navier-Stokes (or Euler) equation, a balance between pressure gradient; Coriolis force; and gravity forces; establishes the link between zonal flows and the induced density perturbation. The gravity term has two contributions, one due to the zonal-flow induced density perturbation and a second due to the dynamic gravity perturbation. If the latter, termed dynamic self gravity (DSG) by \citet{Wicht2020}, is ignored, the balance reduces to the classic thermal wind equation (TWE) which can directly be solved for the density perturbation and hence the gravity perturbation. This approach has been commonly used for modelling the gravity anomalies of Jupiter and Saturn and characterising their deep interior structure \citep{Guillot2018, Kaspi2018, Iess2019, Galanti2019}. Several authors claim that the DSG-term is indeed negligible  \citep{Galanti2017,Kaspi2018}, but this has been disputed by \citet{Zhang2015}, \citet{Kong2017}, and \citet{Wicht2020}. 

Retaining the DSG-related term yields the so called thermo-gravitational wind equation (TGWE), which is harder to handle mathematically and numerically. 
\citet{Zhang2015} reformulate the TGWE into an integro-differential equation for the density perturbation. Their approach is restricted to polytropic interiors with index unity. More recently, \citet{Wicht2020} show that the TGWE can be formulated as an inhomogeneous Helmholtz equation for the gravity potential. The results show that the relative impact of the DSG decreases with the spherical harmonic degree $\ell$ of the gravity perturbation. Being of order one for $\ell=2$, it decreases to $10\%$ at about $\ell=5$ and becomes smaller than $1$\% for $\ell>25$. However, the polytrope provides only a crude approximation of Jupiter's interior. Here we introduce a solution method for the TGWE that is based on formalism of \citet{Wicht2020} but can also handle more realistic interior models that are reflected by a radius-dependent DSG coefficient.

\begin{figure}
\centering
\includegraphics[width=1.\columnwidth]{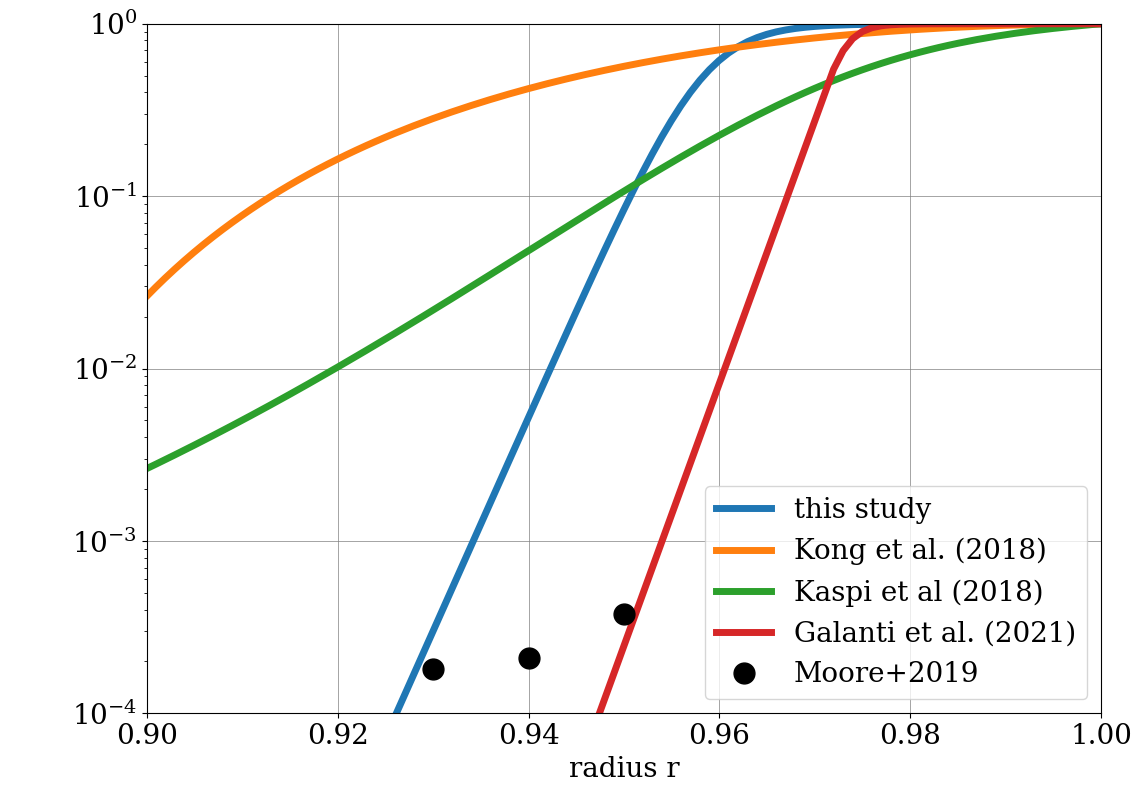}
\caption{Various proposed radial decay functions alongside the expected amplitude of zonal flows that are compliant with the secular variation of Jupiter's magnetic field  \citep{Moore2019}.}
\label{figdecay}
\end{figure}

The discrepancy between existing studies of Jupiter's odd gravity moments is related to differences in the modelling assumptions. \citet{Kaspi2018} explained the first four odd gravity moments of Jupiter by extending the observed zonal wind profile obtained during the perijove \citep{Tollefson2017} along cylinders, inverting the TWE equation, assuming a realistic Jupiter interior model \citep{Guillot03} and a relatively smooth radial decay while ignoring the discontinuity at the equatorial plane (see fig.~\ref{figdecay}, green). This has been challenged by \citet{Kong2017}, arguing that the thermo-gravitational wind equation (TGWE) should be used. The subsequent application to Jupiter by \citet{Kong2018} is based on a polytropic interior and the surface flow measurements from the Cassini mission \citep{Porco2003}. In order to carry this out the surface zonal flow was also modified by putting a cap on the amplitude of the spike in the antisymmetric flow at $\beta=21^\circ$ latitude, and applied a second decay function to smooth out the discontinuity at the equatorial plane (orange profile). 

Both studies favour a rather smooth decay with depth in order to match the observed gravity signal, which implies quite significant flow amplitudes below $0.95\,R$. The radial decay profile of \citet{Kaspi2018} implies  a remaining 10\% flow amplitude at $0.95\,R$, whereas \citet{Kong2018} found more than half of the surface amplitude remaining at this depth. The weak secular variation of Jupiter's observed magnetic field over the past decades, however, can only be explained if the amplitude of the zonal wind at the depth of the horizontal advection of the radial field, i.e. 4\% to 6\% of Jupiter's radius, is as weak as cm/s compared to tens or hundreds of meters per second at the surface \citep{Moore2019} (black dots in the figure). More recently, \citet{Galanti2020} showed that a combined gravito-magnetic analysis favoured a sharper radial profile with a $2000\,$km deep barotropic part and $600\,$km decay region (red profile in fig.~\ref{figdecay}). However, also this analysis neglected the flow discontinuity at the equatorial plane.

In this study we revisit the key model assumptions and properties on which the analyses of odd-degree gravity moments are based and quantify their impact on the solutions. This includes justifying the fundamental equation, the problematic treatment of the equator, various surface zonal flows profiles and different models of Jupiter's interior. 

\section{Theory}
\label{sectheory}
\subsection{Governing equation}
The gravity potential $\varPsi$ is directly related to the density distribution $\rho$ via the Poisson equation
\begin{equation}
    \nabla^2 \varPsi =  4 \pi G \rho  
    \label{eqpoisson} \ ,
\end{equation}
where $G$ is the gravitational constant. The general solution is
\begin{equation}
\varPsi (r,\theta) = 4 \pi G  \int_0^r \int_0^\theta \frac{ \rho(\tilde{r}, \tilde{\theta}) }{\vert\bm{r} - \tilde{\bm{r}} \vert} \tilde{r}^2\, \sin \tilde{\theta} \, d\tilde{r} \, d\tilde{\theta} 
\label{eqgensol}
\end{equation}
and the associated gravity force is $\bm{g} = - \bm{\nabla} \varPsi$. It is useful to separate the external gravity potential into a spherically symmetric part of a non-rotating, solid body and a series of higher order terms originating from density perturbations and  the rotational deformation of the planet:
\begin{equation}
\varPsi =-\frac{G M}{r} \left[1-\sum_{\ell=2}^{\infty} J_\ell \left(  \frac{R}{r} \right)^\ell  P_\ell (\cos \theta) \right] \ ,
\label{eqsolvePsi}
\end{equation}
where $R$ is the equatorial planetary radius, $M$ the total mass and $P_\ell$ are the Legendre polynomials of degree $\ell$. The $\ell=1$ contribution vanishes because the origin of the coordinate system has been chosen to coincide with the center of gravity. The gravity moments $J_\ell$ are given by: 
\begin{equation}
J_\ell = - \frac{2 \pi }{M R^\ell} \int_0^{R} \int_0^\pi\rho (r,\theta) \,  P_\ell(\cos \theta) \,   r^{\ell+2}  \sin \theta \, d\theta \, dr  \ .
\label{eqdefJn}
\end{equation}
While the odd moments contain the signal of the equatorially antisymmetric component of zonal flows, the even moments are dominated by the effects of the rotational deformation of the planet. We therefore focus on the equatorially antisymmetric winds and the respective observed odd gravity moments $J_3$, $J_5$, $J_7$ and $J_9$. 
The general relation between the zonal mass flux and a density anomaly is expressed by the reduced Navier-Stokes equation, that describes the conservation of momentum for a steady, inviscid, non-magnetic, inertia-less flow rotating around the $\hat{\bm{z}}$-axis with a rotation rate $\Omega$. In the co-rotating frame of reference this reads: 
\begin{equation}
2 \bm{\Omega} \times \left( \rho \bm{u} \right) = -\bm{\nabla} p + \rho \bm{\nabla} \varPsi  + \rho \bm{\Omega} \times \bm{\Omega} \times \bm{r} \ ,
\end{equation}
where the terms (from left to right) are the Coriolis force, the pressure gradient, the gravity and the centrifugal force. This leading order force balance applies to the quasi stationary zonal flows in the outer envelope where the electrical conductivity is so low that the Lorentz force can be neglected. The centrifugal force in the leading order force balance represents the rotational deformation and renders the steady background state two-dimensional and non-spherical \citep[e.g.][]{Cao2017}. However, the rotational deformation itself is rather insignificant for the antisymmetric problem \citep{Kong2016} and thus we ignore the centrifugal forces for now.  

Pressure, density and gravity are separated into a hydrostatic background that depends only on radius and a small perturbation, e.g. $\rho = \bs{\rho} (r) + \rho^\prime(r,\theta)$. The first order perturbation equation is then given by:
\begin{equation}
2 \bm{\Omega} \times \left( \bs{\rho} \bm{u} \right) =  - \bm{\nabla} p^\prime + \rho^\prime \bm{\nabla} \bs{\varPsi} + \bs{\rho} \bm{\nabla} \varPsi^\prime \label{eqfirstorder} \ ,
\end{equation}
where the last two terms on the right hand side are gravity force contributions due to a dynamic (i.e. flow-induced) density anomaly ($\rho^\prime \bm{\nabla} \bs{\varPsi}$) and the dynamic self-gravity (DSG,  $\bs{\rho} \bm{\nabla} \varPsi^\prime$) term. In the classic thermal wind approach \citep[e.g.][]{Kaspi2018}, the DSG is neglected. The density anomaly can then simply be found from the thermal wind equation (TWE), which is the azimuthal component of the curl of eq.~\ref{eqfirstorder}:
\begin{equation}
2\Omega \partial_z \left( \bs{\rho} U_\phi \right) = -\frac{1}{r}\bm{\nabla} \bs{\varPsi} \partial_\theta \rho^\prime \ .
\label{eqTWE}
\end{equation}
An integration along latitude and division by the background gravity yields the anomalous density field, which is subsequently used to calculate $J_\ell$ by eq.~\ref{eqdefJn}. 

\citet{Zhang2015} and \citet{Wicht2020} show that the DSG term represents a first order effect and, for example, changes the $J_3$-values by up to $30\%$. Keeping the DSG term in the azimuthal vorticity equation yields the so called thermo-gravitational wind equation (TGWE):
\begin{eqnarray}
2 \Omega \partial_z \left( \bs{\rho} U_\phi \right) &=& \hat{\bm{\phi}} \cdot \left( \bm{\nabla} \times \left(\rho^\prime \bm{\nabla} \bs{\varPsi} + \bs{\rho} \bm{\nabla} {\varPsi^\prime} \right)   \right)  \label{eqcurlfirstorder} \\
 &=&- \frac{1}{r} \frac{d \bs{\varPsi}}{dr} \partial_\theta \rho^\prime   -  \frac{1}{r }\frac{d\bs{\rho}}{dr}\partial_\theta  \varPsi^\prime \ .
\label{eqNSTcomp}
\end{eqnarray}
Replacing $\varPsi^\prime$ by eq.~\ref{eqgensol} and integrating over latitude leads to the integro-differential equation solved  by \citet{Zhang2015}
\begin{eqnarray}
2 \Omega \int_0^\theta \partial_z \left( \bs{\rho} U_\phi(r,\tilde{\theta}) \right)  d \tilde{\theta} = -  \frac{1}{r} \frac{d \bs{\varPsi}}{dr}\rho^\prime \nonumber \\   - \dfrac{1}{r\, }\dfrac{d\bs{\rho}}{dr} \int_0^{r} \int_0^\theta \dfrac{\tilde{r}^2 \rho^\prime}{\vert \bm{r} - \tilde{ \bm{r}}\vert}\sin \tilde{\theta} \, d\tilde{\theta}  d\tilde{r} +C(r)\ .
\label{eqTGWE}
\end{eqnarray}
While the integration function $C(r)$ renders $\rho^\prime$ mathematically non-unique, the gravity moments nevertheless remain unique \citep{Kaspi2010, Zhang2015}. The integration function $C(r)$ would only contribute to the spherical symmetric gravity harmonic which is determined by the total mass and therefore we can set $C(r)$ to zero. Treating eq.~\ref{eqTWE} is mathematically demanding  and has so far only been solved for a simple interior model, i.e. a polytrope of index unity \citep{Zhang2015,Kong2018}.

A potential work-around was devised by \citet{Braginsky1995} in the framework of classic geodynamo theory. It was shown that the DSG term can be absorbed into the so-called {\it effective} variables (density and pressure):
\begin{eqnarray}
p^e &=& p^\prime + \bs{\rho} \varPsi^\prime \\ \rho^e &=&\rho^\prime + \frac{\mu}{4 \pi G} \varPsi^\prime \label{eqeffective} \ .
\end{eqnarray}
The radial function $\mu$ is thereby characterised by the compressibility of the considered medium:
\begin{equation}
\frac{\mu}{4 \pi G} = \bs{\rho} \frac{\partial \bs{\rho}}{\partial \bs{p}}\Bigg\vert_s = \frac{\bs{\rho}}{c_s^2} = \frac{1}{\bs{g}} \frac{d \bs{\rho}}{dr}  \ ,
\label{eqcoeff}
\end{equation} 
where $c_s$ is the sound speed. For a polytropic perfect gas this can be further simplified, 
\begin{equation}
\frac{\mu}{4 \pi G} = \frac{m}{m+1}\frac{\rho_c^{\frac{m+1}{m}}}{p_c}\bs{\rho}^{\frac{m-1}{m}} \ ,
\end{equation}
where $m$ is the polytropic index, $\rho_c$ and $p_c$ are density and pressure at the centre of the planet. They depend on the specific solution of the Lane-Emden equation. Moreover, for a polytropic index unity, $\mu$ is a constant and amounts to $\pi^2/R^2$.

The effective variables, $p^e$ and $\rho^e$, are equal to $p^\prime$ and $\rho^\prime$ reduced by the contribution of the local elevation or depression of the associated equipotential surface thus capturing the effect of the DSG. Using the effective variables and the definition of $\mu$ (eq.~\ref{eqcoeff}), the Navier-Stokes equation (eq.~\ref{eqfirstorder}) simplifies to:
\begin{equation}
 2 \bm{\Omega} \times \left( \bs{\rho} \bm{u} \right) = - \bm{\nabla} p^e + \rho^e \bm{\nabla} \bs{\varPsi} \label{eqeffTWE}  \ .
\end{equation}
Taking the azimuthal component of the curl, integrating along colatitude $\theta$ and replacing $\bm{\nabla} \bs{\varPsi} = \bs{g}$\, , then yields the effective density perturbation:
\begin{equation}
\rho^e (r,\theta) = 2\frac{\Omega r}{ \bs{g}} \int_0^\theta \partial_z \left( \bs{\rho} U_\phi \right) d\tilde{\theta}   \coloneqq \rho^U \ .
\label{eqTWEeff}
\end{equation}
Since this effective density disturbance formulates the zonal flow impact, it has been called $\rho^U$ by \citet{Wicht2020} and we adopt this name here. Note that this is not the true density disturbance $\rho^\prime$ that could serve to calculate $J_\ell$ via eq.~\ref{eqdefJn}. In particular, $\rho^U$ and $\rho^\prime$ are related as defined by eq.~\ref{eqeffective}.

Now we can find an equation for the gravity potential by  replacing $\rho^\prime$ with $\varPsi^\prime$ using eq.~\ref{eqpoisson} in eq.~\ref{eqNSTcomp}, integrating along latitude and making use of eq.~\ref{eqTWEeff}:
\begin{equation}
\nabla^2 \varPsi^\prime + \mu \varPsi^\prime = 4 \pi G \rho^U \ .
\label{eqgentrafo}
\end{equation}

This is a two-dimensional, inhomogeneous PDE of second order, which describes the wind-induced anomalies in the gravity potential of a gas planet \citep{Wicht2020}. The effective density perturbation $\rho^U$ derived from eq.~\ref{eqTWEeff} acts as the source term for this Helmholtz-like equation. Only when $\mu=const$ does this equation become an inhomogeneous Helmholtz equation and can be solved in a semi-analytical way \citep{Wicht2020}. In the more general case where $\mu$ depends on the radius we refer to the numerical methods discussed in sec.~\ref{secnumer}.

It is important to note that the 2nd order differential equation (eq.~\ref{eqgentrafo}) and the integro-differential form of the TGWE (eq.~\ref{eqTGWE}) describe the same physical problem. The main difference is that eq.~\ref{eqgentrafo} solves for $\varPsi^\prime$, while eq.~\ref{eqTGWE} solves for the density anomaly $\rho^\prime$. Eq.~\ref{eqgentrafo} not only directly provides the gravity potential we are interested in, but is also much easier to solve.


\subsection{Geostrophy of the zonal winds}
\label{secadiabat}
The effective variables can be further exploited to show that the flow ($U_\phi$) and not the mass flux ($\bs{\rho} U_\phi$) is geostrophic and should be initially extended along cylinders. To emphasise under which conditions geostrophic winds can be modulated along the rotation axis, we express the density as a function of pressure and entropy:
\begin{equation}
\frac{s^\prime}{c_p} = \frac{c_v}{c_p} \frac{p^e}{\bs{p}} - \frac{\rho^e}{\bs{\rho}} = \frac{c_v}{c_p} \frac{p^\prime}{\bs{p}} - \frac{\rho^\prime}{\bs{\rho}} \ . 
\end{equation}
Dividing the Navier-Stokes equation (eq.~\ref{eqeffTWE}) by the background density and defining a reduced pressure $p^\star = p^e/\bs{\rho}$, eq.~\ref{eqeffTWE} yields:
\begin{align}
    2 \bm{\Omega} \times \bm{u} &= - \frac{1}{\bs{\rho}} \bm{\nabla} p^e + \frac{\rho^e}{\bs{\rho}} \bm{\nabla} \bs{\varPsi} \nonumber\\
    & = -\bm{\nabla} \left( \frac{p^e}{\bs{\rho}}\right) - \frac{p^e}{\bs{\rho}^2} \bm{\nabla} \bs{\rho} + \left(  \frac{c_V}{c_p} \frac{p^e}{\bs{p}} - \frac{s^\prime}{c_p} \right) \bm{\nabla} \bs{\varPsi} \nonumber \\
    & = -\bm{\nabla} \ p^\star - \frac{s^\prime}{c_p} \bm{\nabla} \bs{\varPsi} + \left(  \frac{c_V}{c_p} \frac{\bs{\rho}^2}{\bs{p}}\bm{\nabla} \bs{\varPsi} - \bm{\nabla} \bs{\rho} \right) \frac{p^e}{\bs{\rho}^2 } \nonumber \\
     & \simeq - \bm{\nabla} p^\star - \frac{s^\prime}{c_p} \bm{\nabla} \bs{\varPsi} \label{eqgeoZF} \ .
\end{align}
Note, the term in the brackets scales with the non-adiabaticity of the background state and hence can safely be neglected for a vigorously convecting atmosphere like Jupiter's. This equation highlights that the winds $U_\phi$ and not the mass fluxes are cylindrically invariant. Only if there are sizeable latitudinal gradients in the zonally averaged entropy, are deviations from geostrophy possible, if we restrict the consideration to regions of negligible electrical conductivity. The associated temperature fluctuations required to drive the zonal wind out of its cylindrical invariance can be estimated by azimuthal component of the curl of eq.~\ref{eqgeoZF}:
\begin{equation}
2\Omega \, \partial_z U_\phi = -\frac{g}{c_p r} \partial_\theta s^\prime \ , \end{equation}
where $d\bs{\varPsi}/dr = g$ has been used. Assuming  that the vertical and latitudinal derivative can be approximated with the same length scale (e.g. close to the equator), thus $\partial_z U_\phi \approx \Delta U / \delta $ and $ 1 /r \, \partial_\theta s^\prime \approx \delta s / \delta$. Furthermore, the entropy fluctuations can be (to first order) approximated by temperature fluctuations, thus $ \delta s \approx c_p /\bs{T} \, \delta T$. This then yields 
\begin{equation} 
\delta T \approx \frac{2 \Omega}{g} \bs{T} \, \Delta U  \ .
\end{equation}
For Jupiter $\Omega \approx 1.76 \cdot 10^{-4}\, \mathrm{s^{-1}}$ and $g \approx  25\, \mathrm{m/s^2}$. Then, in order to induce a vertical variation of the wind $\Delta U=10\, \mathrm{m/s}$ at a temperature of $\bs{T} (r=0.95R) = 4\cdot 10^3\, \mathrm{K}$ \citep{Nettelmann2012}, we find  $\delta T \approx 0.5 \,\mathrm{K}$ - an unrealistically high value considering that the temperature fluctuations associated with convective motions are on the order of $10^{-4}\, \mathrm{K}$ \citep[e.g.][]{Jones2007}. These estimates are applicable to the convective part of the atmosphere, i.e. below $p=1\,\mathrm{bar}$.  
In conclusion, whenever convection restricts the degree of non-adiabaticity, the zonal flows $U_\phi$ (and not the mass flux) are cylindrically invariant.

\section{Parametrizing the dynamic density perturbation}
The dynamic density perturbation (eq.~\ref{eqTWEeff}) is governed by the assumed interior structure of the planet via the background density and gravity, $\bs{\rho}$ and $\bs{g}$, respectively, and the $z$-gradient of the zonal mass flux. Theoretical considerations and numerical simulations suggest a geostrophic zonal flow structure for the fast rotation and low viscosity gas planets \citep[e.g.][]{Taylor1917,Dietrich2018, Gastine2021}. 
This means that the flow depends only on the distance 
$s=r\sin{\theta}$ to the rotation axis, where $\theta$ is the colatitude. 
We could then simply downward continue the surface zonal flow $U_0$ along cylinders.

However, the gravity measurements and the secular variation of the Jupiter's magnetic field show that the wind speed must significantly decrease with depth and should be confined to outer 5\% of the planetary radius \citep{Kaspi2018, Moore2019, Galanti2020}. This decrease is commonly parametrized with an additional radial decay function $Q_r$,
\begin{equation}
     U_\phi(r,\theta)=Q_r(r) \;U_o(s)
     \label{eq:geosflow}
\end{equation}  
with $Q_r(R)=1$. The cause for the deviation from geostrophy remains unspecified. Electromagnetic effects have been alluded to. Buoyancy forces arising in a stably stratified region with the assistance of Lorentz forces are a promising mechanism \citep{Christensen2020,Gastine2021}. We start with discussing the interior state, then the different models of the surface zonal profile and finally the calculation of the dynamic density source $\rho^U$.

\subsection{Interior state}
\label{secinterior}
\begin{figure*}
\centering
\includegraphics[width=1.\textwidth]{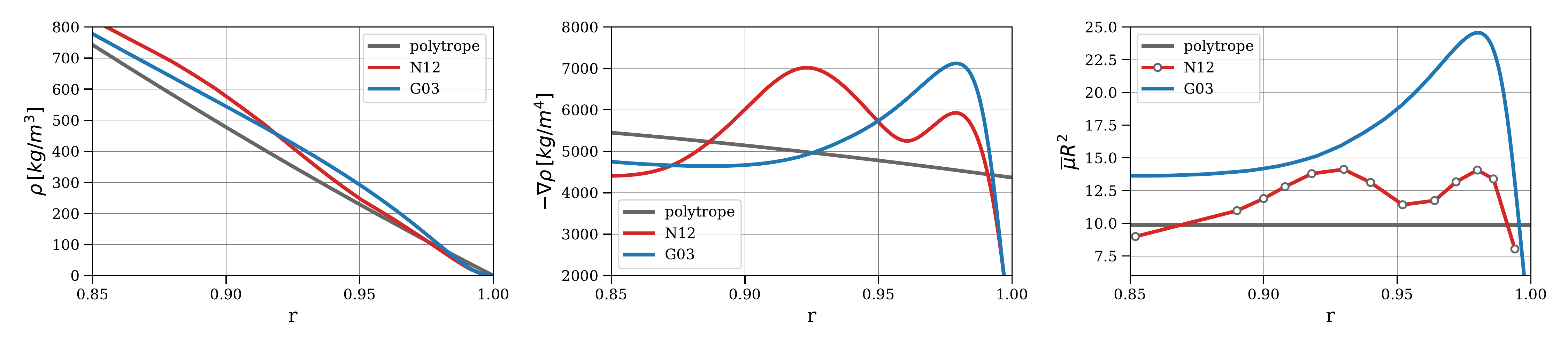}
\caption{Jupiter's interior density $\bs{\rho}$ (left), radial gradient $\nabla{\bs{\rho}}$ and $\mu = \nabla{\bs{\rho}} / g$ in the outer 15\% of radius from two Jupiter interior models in comparison to a polytropic perfect gas with index unity. The interior models based on 3-layer internal structure are adopted from  \citet{Guillot03} (blue) and \citet{Nettelmann2012} (red). The dots on the curve are based in ab-initio calculations of the sound speed ($c_s^2 \propto \mu$) by \citet{French2012}. }
\label{figJinter}
\end{figure*}

The source term (eq.~\ref{eqsourcefin}) and the DSG coefficient, $\mu$, depend on the background density and gravity profile, $\bs{\rho}$ and $\bs{g}$, respectively. We explore their influence via the chosen interior state model by comparing three commonly used models.

A simple model for Jupiter's interior is a polytropic ideal gas of index unity that provides a decent description of the pressure dependence on the density \citep{Hubbard1999}, but not of the ($p-T$)-curve. The background density is then:
\begin{equation}
\bs{\rho}(r) = \bs{\rho}_c \frac{\sin \chi}{\chi} \ , \, \, \text{with} \, \chi= \frac{\pi r}{R} \ ,
\end{equation}
where $\bs{\rho}_c$ is the central density. The associated background gravity is given by
\begin{equation}
\bs{g}(r) = - 4 R  G \bs{\rho}_c \, \frac{\chi \cos \chi - \sin \chi}{\chi^2} 
\end{equation}
and is hence directly proportional to the radial density gradient as discussed in \citet{Zhang2015}. This proportionality ($g\propto d\rho / dr$) is an exclusive property of a polytrope with index unity and implies a constant $\mu = \pi^2/R^2$. This has been exploited by \citet{Zhang2015} and \citet{Wicht2020} to solve the TGWE. The grey profiles in fig.~\ref{figJinter} illustrate the density, its radial gradient and $\mu$ for the polytrope of index $m=1$.  

The first analysis of the Juno gravity data by \citet{Kaspi2018} was based on a Jupiter interior model by \citet{Guillot1994, Guillot03}. This model is henceforth named `G03` and assumes a three layer structure with a helium-depleted, molecular outer hydrogen layer, a helium-enriched metallic inner hydrogen layer and a central dense core. Fig.~\ref{figJinter} shows the density, density gradient and $\mu$ for a setup with interpolated hydrogen EOS, a 1-bar temperature of 165\,K, a core of 4.2 earth masses and heavy element abundance of 33 earth masses \citep{Guillot03}. In comparison to the rather simple polytropic model (grey), the densities are substantially higher for $r<0.975\,R$ and slightly lower above this radius. Thus the density gradient shows a pronounced maximum around $r=0.97\,R$ (fig.~\ref{figJinter}, middle panel). The DSG coefficient ($\mu$) increases with radius and reaches a 2.5 times larger value than for the polytrope. 

Alternatively, we use the more recent calculations from \citet{Nettelmann2012} based on the updated H-REOS2 model (`N12`). The DSG coefficient is related to the sound speed (eq.~\ref{eqcoeff}), which was calculated for the same Jupiter model by \citet{French2012}. For this model (termed J11-8a) the depth of the molecular-metallic phase transition is at $p=8\,$Mbar, whereas in G03 model this happens at shallower 2\,Mbar. As shown in the left panel of fig.~\ref{figJinter} the N12 model yields the highest density for $r<0.92\,R$  and falls between the G03 model and the polytrope at larger radii. The density gradient and hence $\mu$ show two distinct maxima with smaller amplitude than the G03 model. 

\subsection{Jupiter's surface zonal flow}

\label{secaveflow}

\begin{figure*}
\centering
\includegraphics[draft=false,width=0.9\textwidth]{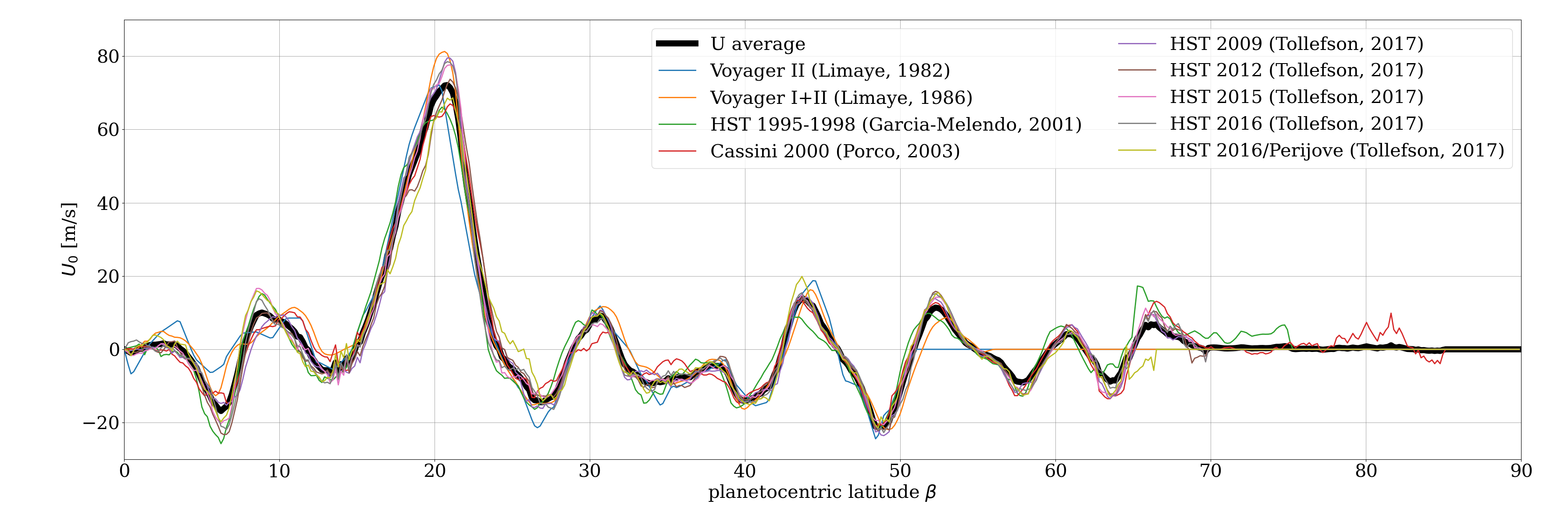}
\caption{
Comparison of different zonal flow models for Jupiter observed either by HST or in-situ by space crafts. Shown is the equatorial antisymmetric flow part relevant for the odd gravity moments. The models are from the Voyager mission \citet{Limaye1982, Limaye1986}, Cassini \citet{Porco2003}, and various Hubble Space Telescope (HST) campaigns \citep{GarciaMelondo2001, Tollefson2017}.}
\label{fig:ZF}
\end{figure*}

The surface flow of Jupiter is deduced from tracking cloud features,
either with the Hubble Space Telescope (HST) or from space crafts. Fig.~\ref{fig:ZF} displays several zonal flow measurements that have been obtained over the last decades. 
The oldest illustrated flows (blue and yellow) are based on  observations by Voyager I and II \citep{Limaye1982,Limaye1986} and represent the flow in 1979. HST monitored the wind structure several times between 1995 and 1998 \citep[green profile,][]{GarciaMelondo2001}. Cassini measurements of the wind profile stem  from late 2000 \citep[red, ][]{Porco2003}. Later, various different HST-based profiles were obtained between 2009 and 206, during the perijove nine of the Juno space craft \citep{Tollefson2017}.
\citet{Kaspi2018} 
and \citet{Galanti2020} use this most recent flow profile for their gravity data analysis, while \citet{Kong2018} prefer the model based on Cassini images from late 2000 to early 2001 presented in \citet{Porco2003}.  

All flow models show the same principle structure but 
also differ in some details. The equatorially antisymmetric
flow contribution shown in fig.~\ref{fig:ZF} is clearly dominated by the prograde jet

which starts at a latitude around $\beta=17^\circ$ and extends to about $\beta=23^\circ$.
The amplitude of this jet varies between $65\,$m/s and 
$75\,$m/s for the different models. 

\citet{Tollefson2017} conclude that the variations
exceed the model errors, at least for some epochs and at some latitudes. Their Lomb-Scargle periodogram analysis reveals dominant variation periods of about $7\,$yr and $14\,$yr. Since the latter is close enough to Jupiter's orbital period of $11.9\,$yr and the former 
to the respective first overtone, the variations may represent seasonal cycles. 
These variation are unlikely to penetrate deeper into Jupiter's atmosphere and therefore play no role for the gravity signal. The arithmetic mean profile, shown as a black line 
in \figref{fig:ZF}, should represent the deeper flows more 
faithfully than a single model. However, the average 
only covers a period of about 36 years, or about three times 
the seasonal cycle, with a small number of `snapshots`. 
Smaller deviations from the time average are thus certainly
conceivable. We mostly rely on the mean wind profile but also explore the impact of using specific snapshots on the gravity signal in sec.~\ref{secflowprof}.

\subsection{Treatment of the equatorial discontintuity}
An obvious problem arises with the equatorially antisymmetric 
contributions to the dynamic density source.  Geostrophically downward continuing
the surface flow in each hemisphere separately yields a discontinuity, 
or step, at the equatorial plane. It can be large in case the geostrophic continuation of the surface flow hits the equatorial plane at a radius where $Q_r$ is significantly larger than zero. This equatorial step seems unphysical but can easily be dealt with mathematically. 

Using the product ansatz (eq.~\ref{eq:geosflow}) in eq.~\ref{eqTWEeff}, we can separate the dynamic density perturbation into two contributions: 
\bel{eq:rhoU2}
\rho^U = - \frac{2\Omega r}{\gb}
\left[ \frac{\partial}{\partial r} \left(\bs{\rho} Q_r\right)
  \int_0^\theta U_o \cos{\hat{\theta}} d \hat{\theta}  
  +\bs{\rho} Q_r \int_0^\theta
\frac{\partial U_o}{\partial z}  d \hat{\theta}\right]
\eep
The second integral contributes only at the equator where the $z$-derivative yields a delta-peak. This can be integrated analytically: 
\begin{equation}
    \rho^U = - \frac{2\Omega r}{\gb}
\left[ \frac{\partial}{\partial r} \left(\bs{\rho} Q_r\right)
  \int_0^\theta U_o \cos{\hat{\theta}} d \hat{\theta}  
  +\frac{\bs{\rho} Q_r}{r}  H(\theta - \pi/2) \Delta U \right] \ ,
\label{eqsourcefin}
\end{equation}
where $H$ is the Heaviside step function and $\Delta U = 2 U_o (\theta \rightarrow \pi/2) $.

\begin{figure}
\centering
\includegraphics[width=1.\columnwidth]{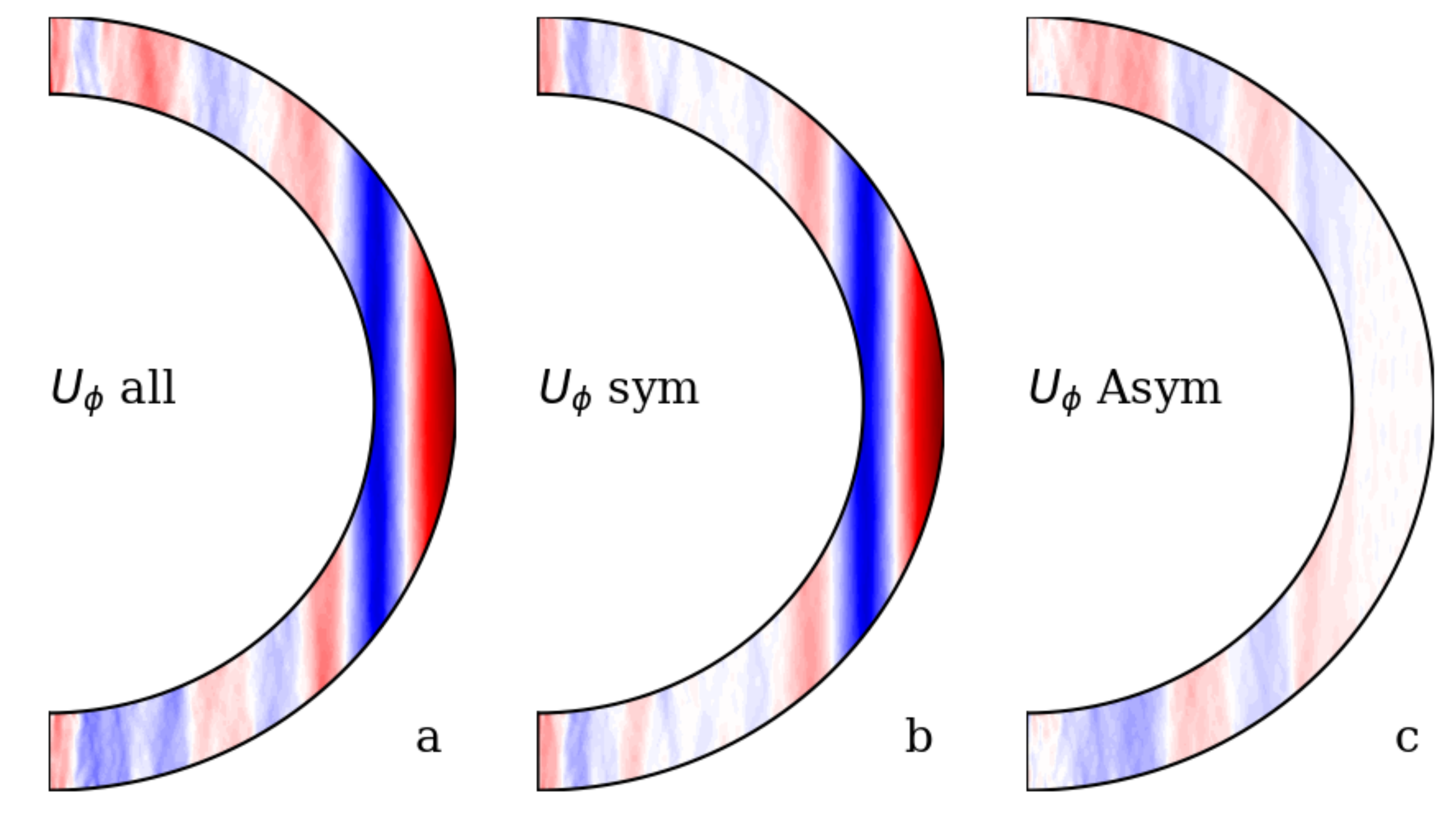}
\caption{Zonal flow from the full 3D numerical model (a) separated into the equatorial symmetric part (b) and the antisymmetric (c) part. The contour levels are identical amongst the panels.}
\label{figWD_AS}
\end{figure}

As we will show below, the equatorial step 
can yield an important contribution to the gravity signal 
which is problematic. 
\citet{Kaspi2018} and \citet{Galanti2020} therefore ignore the respective term in an approach they call `hemispheric`. 
They calculate $\rhoU$ in each hemisphere but dismiss 
the contribution at (or very close to) the equator, which is 
equivalent to evaluating 
\bel{eq:rhoUS}
\rho^U = - \frac{2\Omega r}{\gb}
 \frac{\partial}{\partial r} \left(\bs{\rho} Q_r\right)
  \int_0^\theta U_o \cos{\hat{\theta}} d \hat{\theta}  \ .
\end{equation}
This approach seems to offer a simple fix but
is also inconsistent since the step is an integral part of the chosen flow model.

Alternatively, \citet{Kong2018} use an additional linear $z$-dependence of $U_o$, such that the vertical gradient at the equator is regular and smooth. Their zonal flow model is thus given by 
\bel{eq:QZ}
   U(r,\theta)=Q_r\frac{|z|}{|z_o|}U_o(s)
\eec
where $|z_o|=(R^2-s^2)^{1/2}$ is the local distance from the 
equatorial plane to the surface. 
The equatorial step and the linear $z$-dependence are
the steepest and smoothest end member, respectively, of 
the possible functions that reconcile the 
northern and southern zonal flows.

Here we introduce a novel approach guided by the physical principles of atmospheric dynamics. The generation of deep-reaching zonal flows powered by the statistical correlations of the convective flows (Reynolds stresses) are best captured in  numerical simulations \citep{Heimpel2005, Christensen2002,Dietrich2018,Gastine2021}.  
\Figref{figWD_AS} illustrates the zonal flow 
in a typical simulation of convection in 
a fast rotating spherical shell with an aspect 
ratio of $r_i/R=0.8$ using the MagIC code \citep{Wicht2002, Gastine2012, Schaeffer2013}. 
 For the chosen parameters ($E=3 \cdot 10^{-5}, Pr= 0.25, Ra = 7 \cdot 10^7, N_\rho = 2.3$, definition as in \citet{Gastine2012}), a broad equatorial prograde jet develops, flanked by a deep retrograde jet adjacent to the tangent cylinder (TC). The TC is a virtual cylinder touching the inner boundary at the equator and forms an important dynamical boundary 
in rotating convection. At higher latitudes, i.e. within the TC, alternating jets develop in both hemispheres. These winds show minimal variations in the 
direction of the rotation axis. 
Simulations at even more realistic parameters 
yield solutions with higher degrees of geostrophy 
but are also more numerically demanding 
\citep{Heimpel2016, Gastine2021}.

Panels b) and c) of \figref{figWD_AS} show that the equatorially symmetric flow is dominant outside the TC and that the 
equatorially antisymmetric zonal flow is much weaker
outside than inside the TC. 
Outside of the TC, equatorially antisymmetric flows are 
at odds with geostrophy. Inside the TC, however, 
antisymmetric contributions can be, and indeed
are, highly geostrophic. 
The same dynamical reasons that enforce geostrophy (dominant 
Coriolis force and small viscosity) also 
allow for only weak equatorially antisymmetric flows 
outside the TC.

To account for this physical property of zonal flow in rotating spherical shells, we multiply the antisymmetric surface flow with an additional attenuation function $Q_s$ that reduces the amplitude outside of the latitude
$\beta_\mathrm{TC}$, i.e. where the TC touches the outer boundary:
\begin{equation}
    U^\star_o = U_o(s) \, Q_s \ ,
    \label{eqdefTC}
\end{equation}
with 
\begin{equation}
    Q_s=  \frac{1}{2} \left[ \tanh \left( \frac{\beta-\beta_\mathrm{TC}}{\delta_\beta} \right) - \tanh\left( \frac{\beta+\beta_\mathrm{TC}}{\delta_\beta} \right) \right] +1  \ .
    \label{eqdefQs}
\end{equation}
A second parameter which is introduced here is the width of the latitudinal cut-off, $\delta_\beta$. Both functions, $Q_s (\beta_\mathrm{TC}, \delta_\beta)$ and $Q_r(h, \delta_h)$, define an individual tangent cylinder and hence should be consistent. More details on selecting the best parameter combination for $\beta_\mathrm{TC}$, $\delta_\beta$, $h$ and $\delta_h$ are discussed in sec.~\ref{secmodelling}. When $\beta_\mathrm{TC}$ is sufficiently large and the radial decay function drops rapidly below a certain depth, the resulting flow splits into an independent northern and southern part. Then the equatorial step contribution is identical to zero. The underlying assumption is, that part of the observed zonal flow at cloud level, in particular its antisymmetric parts at low latitude, are shallow. We term cases that apply such an attenuation of antisymmetric surface flows at low latitudes 'TC-models'. 

\subsection{Radial decay function}
The different forms of the  
radial decay function $Q(r)$ that have been suggested to
explain the gravity observations are illustrated in \figref{figdecay}. 
\cite{Kaspi2018} use a combination of 
an exponential decay and a hyperbolic tangent: 
\bea
\label{eq:QK}
Q_r (\alpha, h, \delta_h)& = &\alpha\;\left[\tanh\left(\frac{r -(R-h)}{\delta_h}\right)+1\right]
\large/ \left[\tanh\left(\frac{h}{\delta_h}\right) + 1\right]\nonumber  \\&&
 + \, (1-\alpha)\;
\exp{}\left(\frac{r-R}{h}\right)  \ . 
\eea 
Here $\alpha$ is the weight of the hyperbolic tangent contribution,
$h$ the decay depth and $\delta_h$ the decay width. 
For explaining the gravity observations, 
\cite{Kaspi2018} propose a large 
$\alpha=0.92$, a relatively large $\delta_h =1570\,$km and a depth 
of $h =1803\,$km. This results in a smoothly decaying radial profile (green) illustrated in \figref{figdecay}.

\citet{Kong2018}, on the other hand, used an inverse Gauss profile
\begin{equation}
    Q_r(h,H) = \exp \left[\frac{1}{h} \left( 1 - \frac{H^2}{H^2-(1-r)^2}\right) \right] \ ,
\end{equation}
for $r>H$, and $Q_r = 0.0$ for $r \le H$. \citet{Kong2018} report that the observations are best 
matched with a combination of $H = 10\,484\,$km and 
$\delta_h =15\,377\,$km. 
\figref{figdecay} shows that the respective profile (orange) 
decays also rather smoothly. 

Both the solution suggested by \citet{Kaspi2018} and by \citet{Kong2018} 
are not compatible with the magnetic observations \citep{Moore2019}. Furthermore, vigorous convection provides an almost adiabatic environment and hence the decay functions should be rather flat with a sharp drop at greater depth (see also  sec.~\ref{secadiabat}).
Realizing this, \citet{Galanti2020} proposed 
a steeper decaying alternative with 
a hyperbolic tangent outer, and an exponential inner branch:
\begin{eqnarray}
Q_r= \begin{cases}
\alpha\,\tanh\left(\frac{r-r_T}{\delta_h}\right)
 \,\large/\,\tanh\left(\frac{h}{\delta_h}\right)
+1-\alpha &\mbox{for}\; r\ge r_T  \\
  (1-\alpha)\,\exp{}\left(\frac{r-r_T}{\delta^\prime_h}\right)
  & \mbox{for}\; r<r_T \  .
\end{cases}
\label{eq:QG}
\end{eqnarray}

They suggest $\alpha=0.45$, 
$h =2002\,$km, $r_T/R = 0.972$ and $\delta_h =\delta_h^\prime =204\,$km in
conjunction with an optimised surface flow profile to
explain the gravity observations. The radial profile drops almost faster than the magnetic constraints require (see fig.\ref{figdecay}).

We adopt the pure hyperbolic tangent profile:
\bel{eq:Qtanh}
Q_r(h, \delta_h) = \left[\tanh\left(\frac{r-(R-h)}{\delta_h}\right)+1\right]
\large/ \left[\tanh\left(\frac{h}{\delta_h}\right) + 1\right]
\eep 
For $\delta_h = 200\,$km and $h = 2000\,$km the profile would be virtually identical to the one suggested by \citet{Galanti2020}. However, our analysis favours a substantially deeper flow with $h =3000\,$km and $\delta_h = 500\,$km. Fig.~\ref{figdecay} shows the respective decay function in blue.

\section{numerical method}
\label{secnumer}
The semi-analytical method for solving the TGWE (eq.~\ref{eqgentrafo}) developed by \citet{Wicht2020} is restricted to the case of a constant $\mu$. However, as shown in fig.~\ref{figJinter}, $\mu(r)$ varies strongly and reaches much higher values than $\pi^2/R^2$, particularly in the outer 10\% of Jupiter's radius where the flow-induced gravity moments originate from. 
We have therefore developed a numerical method that can also handle radial variations in $\mu$.

We first formulate the dynamic density source $\rho^U(r, \theta)$ via  eq.~\ref{eqsourcefin} in spatial space by choosing a radial decay function $Q_r$, an interior state model setting $\bs{\rho}$, $\bs{g}$ and $\mu$, and a surface zonal flow profile $U_o$, that is cylindrically downward continued. 

We then expand the latitudinal dependence of the gravity potential perturbation $\varPsi^\prime$ and the dynamic density $\bs{\rho}^U$ in Legendre polynomials, e.g. for the former this is given by
\begin{equation}
\varPsi^\prime( r,\theta) = \sum_{\ell=0}^{L}  \varPsi_{\ell}^\prime(r) P_\ell(\cos \theta) \ , \label{eqSpecLeg}
\end{equation}
with
\begin{equation}
    \varPsi_\ell^\prime(r) = \int_0^\pi \varPsi^\prime(r,\theta) P_\ell (\cos \theta) \sin \theta d\theta \ .
\label{eqlegtrans}
\end{equation}
Since the background state, and hence $\mu$, are only a function of radius, the TGWE (eq.~\ref{eqgentrafo}) decouples for each degree $\ell$, and we are then left with a set of radial ODEs, one for each spherical harmonic degree $\ell$: 
\begin{equation}
\left( \partial_r^2 + \frac{2}{r} \partial_r  -  \frac{\ell(\ell+1)}{r^2} +\mu \right) \varPsi_\ell^\prime(r) = 4 \pi G \rho_\ell^U(r)  \ .
\label{eqanaHelm}
\end{equation}
At the outer boundary the potential must match to the solution of a source free region, i.e. $\nabla^2 \varPsi_{pot} = 0$, with the characteristic radial dependence $r^{-(\ell+1)}$. This leads to the mixed outer boundary condition:
\begin{equation}
    \varPsi_\ell^\prime (R) = -\frac{R}{\ell+1} \partial_r \label{eqdefBCbes} \varPsi_\ell^\prime (R) \ ,
\end{equation} 
and $\varPsi^\prime_\ell(0) = 0$ at the inner boundary.

Our numerical method relies on an expansion in the modified spherical Bessel functions introduced by \citet{Wicht2020}. These are radial eigenfunctions of the homogeneous Helmholtz equation. For each spherical harmonic degree $\ell$ we construct a set of N normalised orthogonal functions $j^\star_{\ell n}(k_{\ell n} r)$. The radial scales, $k_{\ell n}$, are determined by finding the first N radii (starting at the origin), at which $j^\star_{\ell n}(k_{\ell n} r)$ fulfils the boundary condition (eq.~\ref{eqdefBCbes}). The radial functions $\varPsi^\prime_\ell(r)$ are expanded in $j^\star_{\ell n}$:
\begin{equation}
    \varPsi_\ell^\prime(r) = \sum_{n=1}^{N} \varPsi^\prime_{\ell n}  j^\star_{\ell n} (k_{\ell n} r) \ , 
    \label{eqspecBes}
\end{equation}
where the expansion coefficients $\varPsi^\prime_{\ell n}$ are  given by
\begin{equation}
    \varPsi^\prime_{\ell n} = \int_0^{R} \varPsi_\ell^\prime(r) \, j^\star_{\ell n} (k_{\ell n} r) \, r^2 dr \ .
\end{equation}
Since the modified spherical Bessel functions are eigenfunctions of the Laplace operator, eq.~\ref{eqanaHelm} transforms into a set of linear algebraic equations for the expansion coefficients $\varPsi^\prime_{\ell n}$:
\begin{equation}
   \sum_{n=1}^N \left( \mu(r) - k_{\ell n}^2\right) \varPsi^\prime_{\ell n} j^\star_{\ell n} (k r) = 4 \pi G \rho^U_{\ell} (r) \label{eqvecform}\ .
\end{equation}
Eq.~\ref{eqvecform} defines a coupled set of algebraic equations and solved by using a matrix formalism. The source term, ${\rho}_\ell^U$, on the RHS is discretized along $N$ radial grid points, $r_i$. The solution of the matrix equation, $\varPsi^\prime_{\ell n}$, contains the Bessel function expansion coefficients of the gravity field perturbation. We thus introduce a square matrix $\bm{\mathcal{H}}_\ell$ defined by
 \begin{equation}
     \mathcal{H}_{\ell n \, i}  = \left( \mu(r_i) - k_{\ell n}^2 \right) j^\star_{\ell n}(k_{\ell n} r_i) \ ,
\end{equation}
where $n,i \in [1,N]$. Then the TGWE with radially varying DSG coefficient can be written in the symbolic matrix form
 \begin{equation}
     \bm{\mathcal{H}}_\ell \bm{\varPsi}^\prime_{\ell n} = 4 \pi G \bm{\rho}_\ell^U  \ ,
     \label{eqmatrix}
 \end{equation}
 where $\bm{\rho}_\ell^U$ represents the source vector per degree along all radial grid points and $\bm{\varPsi}^\prime_{\ell n}$ is the solution vector containing the expansion coefficients. The matrix equation is solved via LU factorization. For dealiasing, we thus ignore the highest 10\% of the Bessel coefficients. We use an equidistant radial mesh with up to $N=1024$ grid points, Gauss-Legendre points along latitude ($N_\theta =4096$) for an accurate Legendre expansion (eq.~\ref{eqlegtrans}) and solve for the first four odd spherical harmonic degrees ($\ell=3, 5, 7, 9$). 

 Having obtained the solution $\varPsi^\prime_{\ell n}$ in Bessel space, eq.~\ref{eqspecBes} yields the radial representation $\varPsi^\prime_\ell$. The gravity moments are then simply given by
\begin{equation}
    J_\ell = \frac{R}{G M} \frac{2\ell+1}{2} \, \varPsi^\prime_\ell(R) \ ,
\end{equation}
If required, the density perturbation can be calculated by solving $\nabla^2 \varPsi^\prime = 4 \pi G \rho^\prime$. 

To test our method, we compare the solution for constant $\mu$ with results using the semi-analytical method introduced by \citet{Wicht2020}. To verify our method with radius-dependent $\mu$, we solve eq.~\ref{eqanaHelm} with an independent solver based on the shooting method. This is a standard tool for solving initial value problems and can readily be applied here, making use of a variable transformation to account for the asymptotic behaviour at the origin when applying this method. 

\section{Modelling the gravity perturbations}
\label{secmodelling}
\begin{figure}
\centering
\includegraphics[width=1.0\columnwidth]{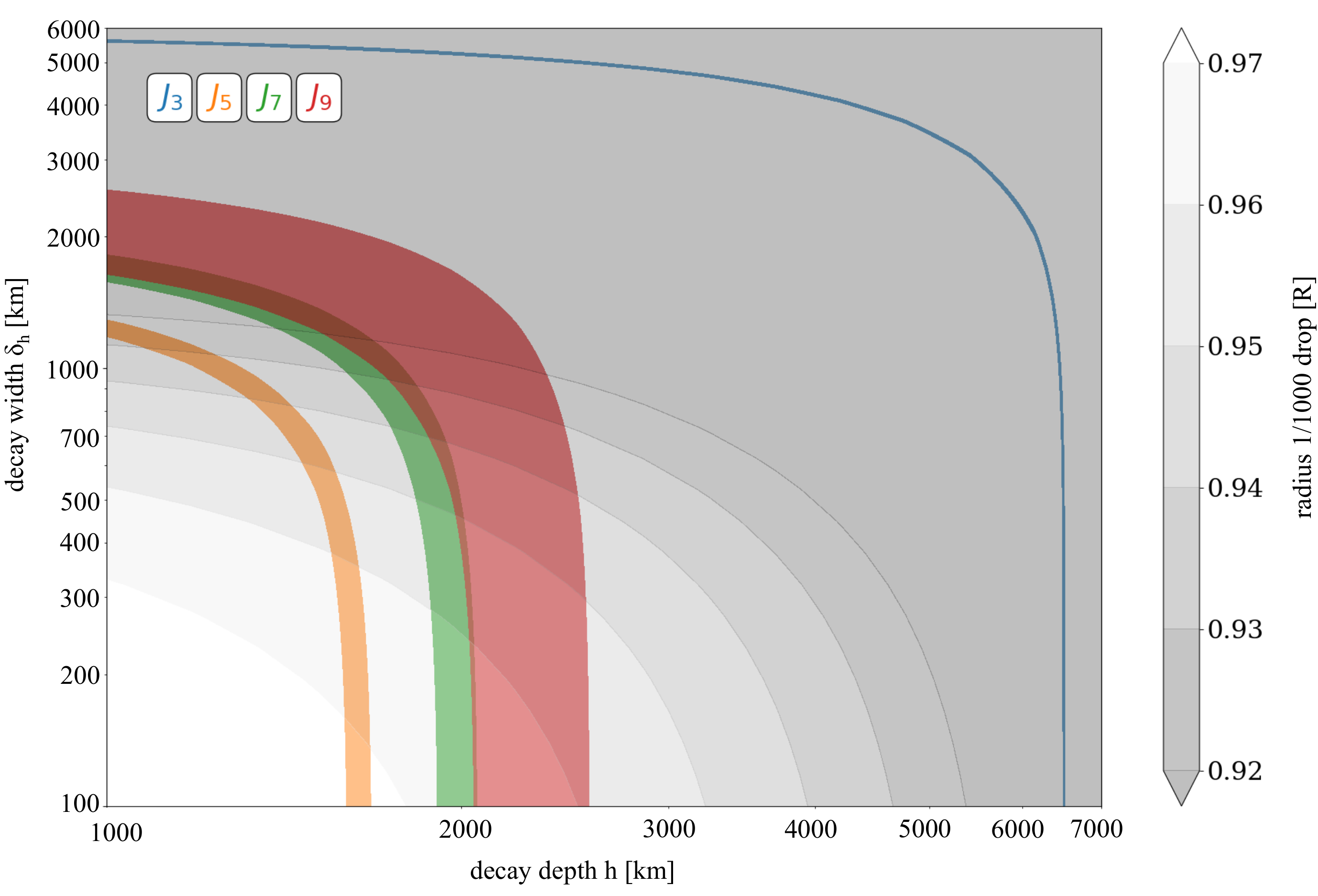}
\caption{Trade-off between decay depth $h$ and decay width $\delta_h$. The coloured contours indicate combinations, where individual gravity moments are compatible within the uncertainties with the measurements ($J_3$ blue, $J_5$ orange, $J_7$ green and $J_9$ red), while the grey shaded areas indicate a flow drop of three orders of magnitude at various depths as required by Jupiter's secular variation \citep{Moore2019}.}
\label{figdeg}
\end{figure}

\begin{figure*}
\centering
\includegraphics[width=1.0\textwidth]{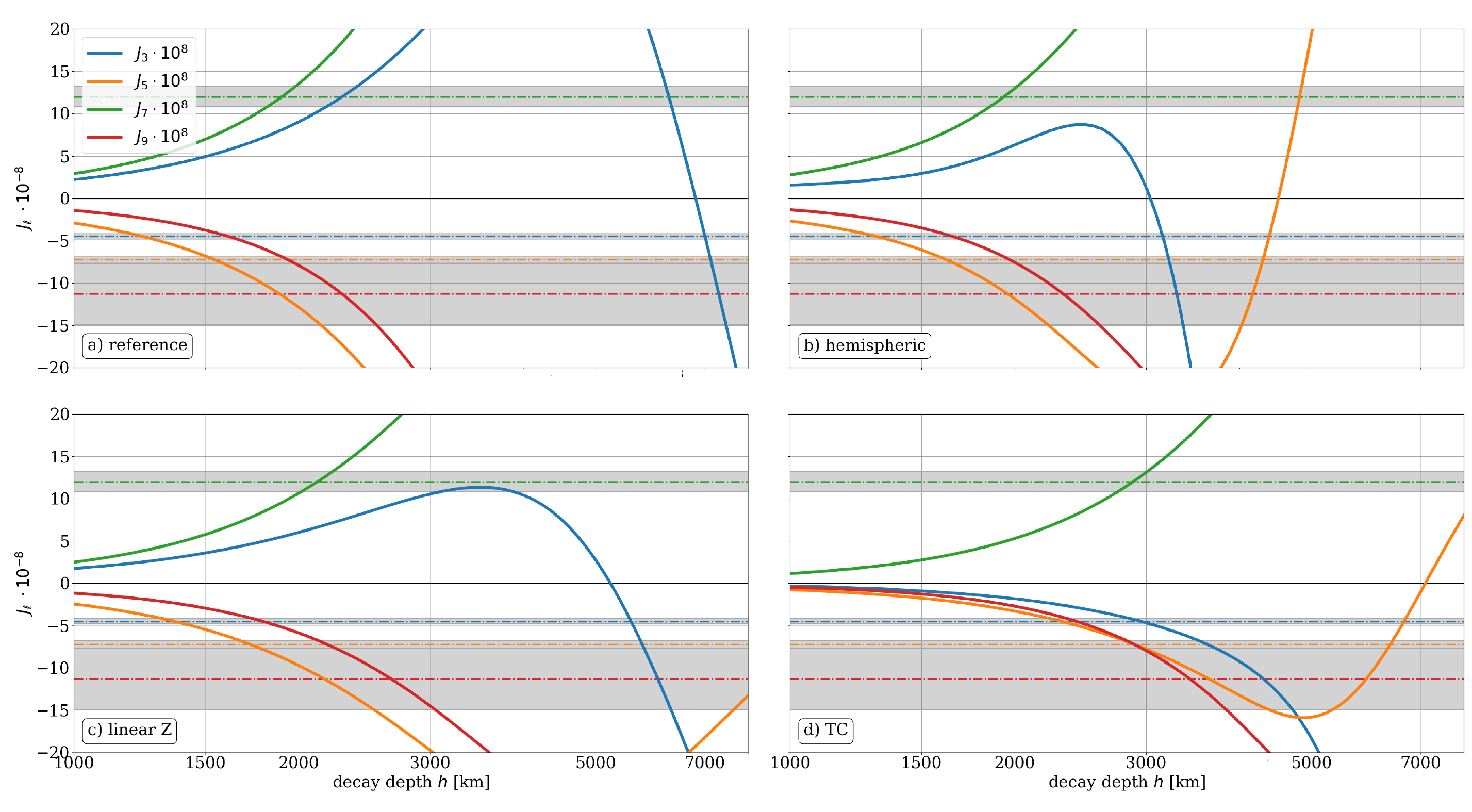}
\caption{Odd gravity moments (degree $\ell=3,5,7,9$ in blue, orange, green and red) as function of the decay depth alongside the measurements (horizontal lines) and $3\sigma$ uncertainties \citep{Durante2020}. The panel a) represents the reference model using the full surface flow, a geostrophic downward continuation and inlcuding the equatorial step, whereas in b) the equatorial step is ignored. In c) the additional linear $z$-decay is applied, and the d) utilises a zero flow outside TC. More details in the text or in tab.~\ref{tabmodres}. }
\label{figmodel}
\end{figure*}

We start by exploring the challenges of 
modelling the odd gravity perturbations 
for a reference model that combines
the interior model by \citet{Guillot03} with
the average flow introduced in sec.~\ref{secaveflow} and employs the TGWE method (eq.~\ref{eqanaHelm} or \ref{eqvecform}). 
Furthermore  the dynamic density source is calculated via eq.~\ref{eqsourcefin} and thus we assume a hemispherically geostrophic flow, keep the equatorial step and apply the hyperbolic tangent radial decay profile for $Q_r$ in accordance with eq.~\ref{eq:Qtanh}.

\subsection{Reference model and equatorial treatment}
Fig.~\ref{figdeg} shows an attempt to model the 
observed gravity harmonics by varying the two 
parameters in $Q_r$, the decay depth $h$ and the decay width $\delta_h$.  A latitude-dependent attenuation in the form of eq.~\ref{eq:QZ} or eq.~\ref{eqdefTC} is not applied at this point.  To assess the quality of the modelled gravity moments, we compare them individually for each degree $\ell$ to the observations rather than minimising an $\ell$-independent, global cost function \citep{Kaspi2018}.
Semi-transparent, coloured, regions indicate the parameter combinations
where the respective gravity harmonic stays within $3\sigma$ 
of the observations \citep{Durante2020}.  There is a trade-off between both parameters since increasing either boosts the impact of 
deeper regions. Unfortunately, there is no parameter combination
where all stripes overlap, i.e. no combination of $h$ and $\delta_h$ that would explain all four odd harmonics with one single decay function. Gravity harmonic $J_3$ seems to be particularly problematic as it always maintains some distance from the other three, since to match it requires particularly deep sources.

The grey background contours show at which radius the radial decay function $Q_r$ would have decayed by three orders of magnitude. The SV constraint by \citet{Moore2019} suggests that this should happen somewhere between 
$r = 0.93\,R$ and $r = 0.96\,R$. Modelling the observed $J_3$ always requires deeper sources that are incompatible with this constraint. For the smooth decay functions suggested by \citet{Kaspi2018} or \citet{Kong2018} the three orders of magnitude drop lies far below $r=0.90\,R$ (see also fig.~\ref{figdecay}).

Numerical simulations and theoretical consideration suggest that the flow remains geostrophic until a stably stratified layer, Lorentz forces or a combination of both, drastically quench the amplitude over a few hundred kilometres \citep{Christensen2020,Wicht2020b}. We therefore restrict our analysis to $\delta_h=500\;$km in the following, a value that represents the scale height of the electrical conductivity in the outer atmosphere of Jupiter  quite well \citep{French2012, Wicht2019} and will be further validated from the results of the TC model (see sec.~\ref{secTCmodel}).

\Figref{figmodel} illustrates how the 
gravity harmonics change when varying $h$ while keeping 
$\delta_h=500\,$km fixed. For the reference model (Fig.~\ref{figmode}, a), the harmonics $J_5$, $J_7$, and $J_9$ all 
agree with the respective observations for $h$ values between $1500\,$km and $2300\,$km. However, $J_3$ requires much deeper flows with $h \approx 6500\,$km. Tab.\ref{tabmodres} gives the values for $h$ and the respective uncertainties for all considered models. We use the spherical harmonic degree $\ell$ as an index to denote the different values of $h_\ell$ required to explain the different observations. For a successful model, all four $h$-values must agree within the errors. This is clearly not the case for our reference model. Only for degree $\ell=7$ and $\ell=9$ the decay depths match within the uncertainties at $h \approx 2000\,$km. For reasonable values of $h$ between $2000\,$km and $4000\,$km the modelled gravity moment $J_3$ is of the wrong sign, whereas $J_5$ is much larger than the observed value in that range of $h$.

Panel b) in fig.~\ref{figmodel} illustrates that the situation improves when we follow the approach by \citet{Kaspi2018} and ignore the contribution due to the 
equatorial step in eq.~\ref{eqsourcefin}. While the values of $h_5$, $h_7$ and $h_9$ remain nearly unchanged, $h_3$ decreases by half to about $3200\,$km (see also tab.~\ref{tabmodres}).  This shows the large impact of the equatorial discontinuity, in particular on degree $\ell=3$. 
If in addition to this we assume the smoothly decaying radial function $Q_r$ suggested by \citet{Kaspi2018}, ignore the DSG term and utilise the perijove surface flow model, we can largely reproduce their results and find a reasonably good agreement of the decay depths across the degrees.

Panel c) of fig.~\ref{figmodel}) shows the results when using the additional linear z-dependence suggested by \citet{Kong2018}. This  leads to shallower winds closer to the equator, a smooth vertical gradient and thus no equatorial step. Again, $h_3$ decreases significantly, but not as much as for the model that ignores the equatorial step. The other harmonics are somewhat more affected, such that all $h_\ell$ increase by roughly $10\%$ with respect to the reference model (tab.~\ref{tabmodres}). Though this additional modification of the vertical flow structure avoids the equatorial discontinuity, it is hard to justify physically.
Again we would have to also adopt the other model ingredients (TGWE model with polytropic interior, Cassini flow with capped spike, inverse Gauss decay profile) to reproduce the results by \citet{Kong2018} and match all harmonics.

\subsection{TC model}
\label{secTCmodel}
\begin{figure}
\centering
\includegraphics[width=0.45\textwidth]{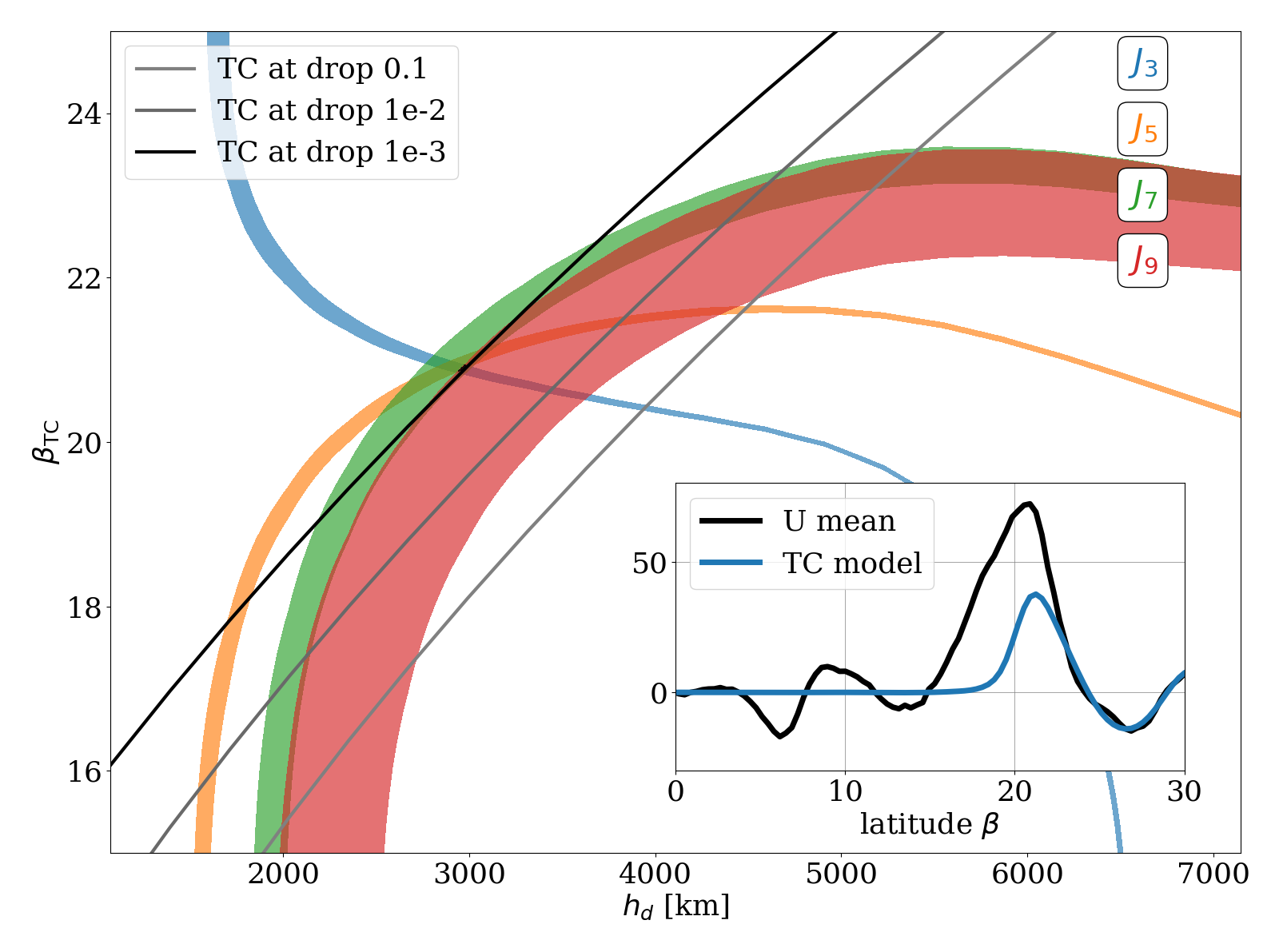}
\caption{Two-dimensional parameter scan with respect to the decay depth $h$ and the assumed TC angle $\beta_{TC}$. The coloured isocontours indicate where the modelled gravity moments agree with the measurements and its uncertainties. The dark lines show the surface latitude of a TC attached to tenfold drop-offs of the associated decay function. The optimal parameter choice is $\beta_\mathrm{TC}=20.9^\circ$ and $h=2975\,$km. The small inset shows the resulting surface flow profile (blue) in comparison with the full flow (black).}
\label{fig2Dcut}
\end{figure}

Finally we explore our TC model where the flow amplitude outside of an assumed tangent cylinder is reduced by applying an additional attenuation function (eq.~\ref{eqdefQs}). The main parameters to examine are the latitude of the TC, $\beta_{TC}$, and the decay depth $h$ (fig.~\ref{fig2Dcut}). Like in fig.~\ref{figdeg} the transparent stripes show where individual gravity harmonics are in agreement with the observed values \citep{Durante2020}. The TC decay width and the radial decay width are fixed to $\delta_\beta=2^\circ$ and
$\delta=500\,$km respectively. For small $\beta_{TC}$ the solution is not affected, implying that the antisymmetric zonal flow between a latitude of $\pm 15 ^\circ$ is irrelevant for the gravity signal. This changes once the TC angle is increased beyond $\beta_{TC}=15^\circ$ since we start reducing the amplitude of the dominant prograde jet at about $\beta = 21^\circ$ latitude. The required decay depth $h_\ell$ decreases for $\ell=3$ but increases for 
$\ell=5$, $7$, and $9$. At $h=2975\,$km and $\beta_{TC}=20.9^\circ$ the model can finally explain all observed gravity harmonics (see also tab.~\ref{tabmodres}). The corresponding surface flow of the TC model is shown in the inset of fig.~\ref{fig2Dcut} (blue profile) indicating that the dominant jet around $20^\circ$ is thinner than in the reference model (black) and reduced to a peak amplitude of $35\,$m/s. Also \citet{Kong2018} found it necessary to reduce the amplitude of the prograde jet.

Reducing the flow in an equatorial latitude band reduces the impact of the equatorial step on the density anomaly. It even vanishes when the TC is positioned at sufficiently large latitudes such that the flow is split into two independent parts. This must be consistent with the assumed radial decay function $Q_r$: a deeper-reaching flow (larger $h$ or $\delta_h$) requires a larger $\beta_\mathrm{TC}$ to separate the northern and southern hemisphere. Thus, a geometric relation between radial decay function $Q_r$ and the width of the equatorial cut-off $\beta_\mathrm{TC}$ can be formulated. The dark curves in fig.~\ref{fig2Dcut} indicate the position of a TC attached to various depths; i.e. where the associated drop-off equals $10^{-1}$ (light grey), $10^{-2}$ (dark grey) or $10^{-3}$ (black). Interestingly, the black line exactly hits the intersection point of the coloured isocontours. This means that the TC defined by reducing the flow amplitude in an equatorial band of $\pm 20.9^\circ$ coincides with the TC defined by the $10^{-3}$-drop-off of the radial decay function $Q_r$. Thus this marks the point in the ($\beta_\mathrm{TC}$-$h$)-parameter space, where the flow is just split into two separate hemispheric flows at the minimum possible  $\beta_\mathrm{TC}$.

Moreover, the other two parameters, $\delta_h$ and $\delta_\beta$, can be constrained with this result. Higher values of $\delta_\beta$ do not lead to an intersection of the individual gravity solutions (colours in fig.~\ref{fig2Dcut}), whereas larger or smaller values of $\delta_h$ shift the black line out of the intersection point.

Consequently, the TC gravity model not only explains all gravity measurements; it is also independent of the handling of the equatorial step since the geostrophic extension of the antisymmetric flows in either hemisphere do not reach the equatorial plane.

Fig.~\ref{figmodel} d) illustrates how the gravity harmonics change for $\beta_{TC}=20.9^\circ$ when increasing $h$. 
The TC scenario  most strongly affects $J_3$ which now remains negative for all $h$. However, the other harmonics are more significantly changed than by the different approaches to treat the equatorial discontinuity illustrated in fig.~\ref{figmodel}. The subsequent $h_3$ is much smaller than in other models, whereas the other $h$ increase drastically. All $h_\ell$ are in agreement within the uncertainties at approximately $h = 2975\,$km (see also tab.~\ref{tabmodres}). 

\subsection{High sensitivity of $J_3$}
\begin{figure*}
\centering
\includegraphics[width=0.9\textwidth]{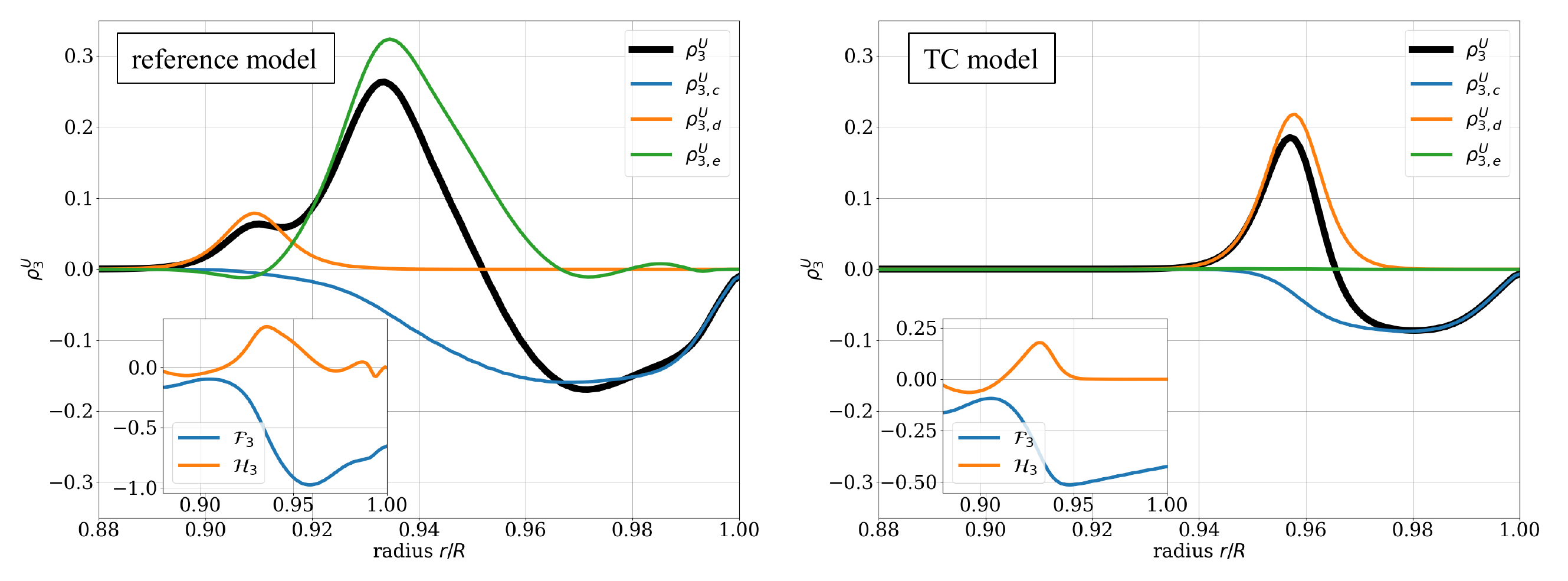}
\caption{Radial profiles of the dynamic density source for degree $\ell=3$, separated into three parts (constant, decay and equatorial step part, see eq.~\ref{eqsourcepart}). The left panel shows the reference model with $h=6541\,$km, the right one the TC model with $h=2975\,$km. The small insets give the radial functions $\mathcal{F}_3$ and $\mathcal{H}_3$. }
\label{figrho3part}
\end{figure*}

Evidently the $J_3$ gravity moment is highly sensitive to different models for the equatorial treatment. To understand this, we analyse the respective source - the dynamic density perturbation $\rho^U_3$ - in more detail. Separating the first term in eq.~\ref{eqsourcefin} into two and transforming in Legendre space yields three source contributions for $\ell=3$:

\begin{equation}
    \rho_3^U = \underbrace{- \frac{2\Omega r}{\gb}
 Q \left( \partial_r \bs{\rho} \right) \, \mathcal{F}_3}_{\rho^U_{3,c}} \underbrace{ - \frac{2\Omega r}{\gb}
 \bs{\rho} \left( \partial_r Q \right) \, \mathcal{F}_3}_{\rho^U_{3,d}} \underbrace{ - \frac{2 \Omega}{g} \rho Q \mathcal{H}_3 }_{\rho^U_{3,e}} \ ,
 \label{eqsourcepart}
\end{equation}
where $\mathcal{F}_3$ and $\mathcal{H}_3$ exhibit the Legendre-transformations of the integrated flow and the step contribution: 
\begin{eqnarray}
    \mathcal{F}_3 &=&\int_0^\pi P_3 \int_0^\theta U_o \cos{\hat{\theta}} d \hat{\theta} \, \sin \theta d\theta \\
  \mathcal{H}_3 &=&\Delta U   \, \int_0^\pi P_3 \, H(\theta - \pi/2) \, \sin \theta d\theta   \ .
\end{eqnarray}
The first part, $\rho^U_{3,c}$ mainly represents the geostrophic part of the flow (where $Q_r$ is constant), the second term, $\rho^U_{3,d}$ is dominated by the decay region and the last, $\rho^U_{3,e}$ shows the influence of the equatorial step. In fig.~\ref{figrho3part} the left panel shows the resulting profiles for the reference model with $h=6541\,$km, i.e. where the gravity model reproduces the observed $J_3$. The right panel displays the TC model with $h=2975\,$km. The radial functions $\mathcal{F}_3$ and $\mathcal{H}_3$ are plotted in the insets of the respective model.

For the reference model, the first contribution $\rho^U_{c}$ (blue) is proportional to the radial derivative of $\bs{\rho}$ and is therefore negative since $\mathcal{F}_3$ is also negative. It dominates at shallow depths $r>0.95\,R$. The second source contribution $\rho^U_{d}$ (orange) is proportional to the radial derivative of the depth profile $Q_r$ and therefore positive. It is only significant between $0.90\,R$ and $0.92\,R$ where 
the large $h$-value positions the steep decrease in $Q_r$. The flow amplitude represented by $\mathcal{F}_3$ is rather small at this depth hence the small positive bump in $\rho^U_{d}$.

The third contribution, $\rho^U_{e}$ (green) is mostly positive (due to $\mathcal{H}_3$) and thus has the wrong sign for explaining the negative $J_3$. For a successful model, the negative $\rho^U_c$ has to overcompensate $\rho^U_{d}$ and $\rho^U_{e}$ which only succeeds when $\rho^U_c$ reaches deep enough. Both $\rho^U_c$ and $\rho^U_e$ individually would yield much larger $J_3$ amplitudes
than required. Only the delicate balance
between the three contribution yields the correct sign and amplitude. This explains the sensitivity of $J_3$ to the reference model setup.  

The source analysis also readily explains why ignoring $\rho^U_e$ in the hemispherical model (panel b) of fig.~\ref{figmodel}) allows one to explain $J_3$ with a much smaller value of $h_3$. The linear $z$-dependence assumed for panel c) of fig.\ref{figmodel} only decreases the amplitude of $\rho^U_e$ and is therefore less effective. 

For the TC model (right panel of fig~\ref{figrho3part}) the equatorial step contribution, $\rho^U_{3,e}$ is identical to zero validating the results from the previous chapter (see also fig.~\ref{figmodel}, d). The density anomaly entirely comprises of a shallow, negative $\rho^U_c$ and deep-rooted positive decay part $\rho^U_d$. Evidently, $\rho^U_d$ is larger because $\mathcal{F}_3$ is larger at the depth of the decay (close to $0.96\,R$). 

\subsection{Influence of dynamic self-gravity}
We have mentioned above that some authors ignore the dynamic self-gravity or DSG-term \citep{Kaspi2018, Galanti2020}, which allows them to solve the TWE rather than the TGWE equation. We model TWE-based gravity moments by setting $\mu=0$ in eq.~\ref{eqgentrafo}. Further, in order to isolate the importance of a realistic radial $\mu$-profile we model the constant coefficient TGWE equation used by \citet{Zhang2015, Kong2018, Wicht2020} by calculating gravity moments with a constant $\mu=\pi^2/R^2$. 

Tab.~\ref{tabmodres} lists the decay depth $h_\ell$ that is required to to match the observed gravity signal. Ignoring the DSG term (TWE model, or $\mu=0$) shows a reduction of $h_3$ by almost $1000\, \mathrm{km}$ (or 15\%) and an increase of $h_5$ by $170\,$km (10\%). In general, the absolute difference shrinks with the degree $\ell$. The model with $\mu=\pi^2/R^2$ shows an intermediate behaviour indicating that a more realistic radial profile of $\mu$ is an important part of the model.

\begin{figure}
\centering
\includegraphics[width=1.\columnwidth]{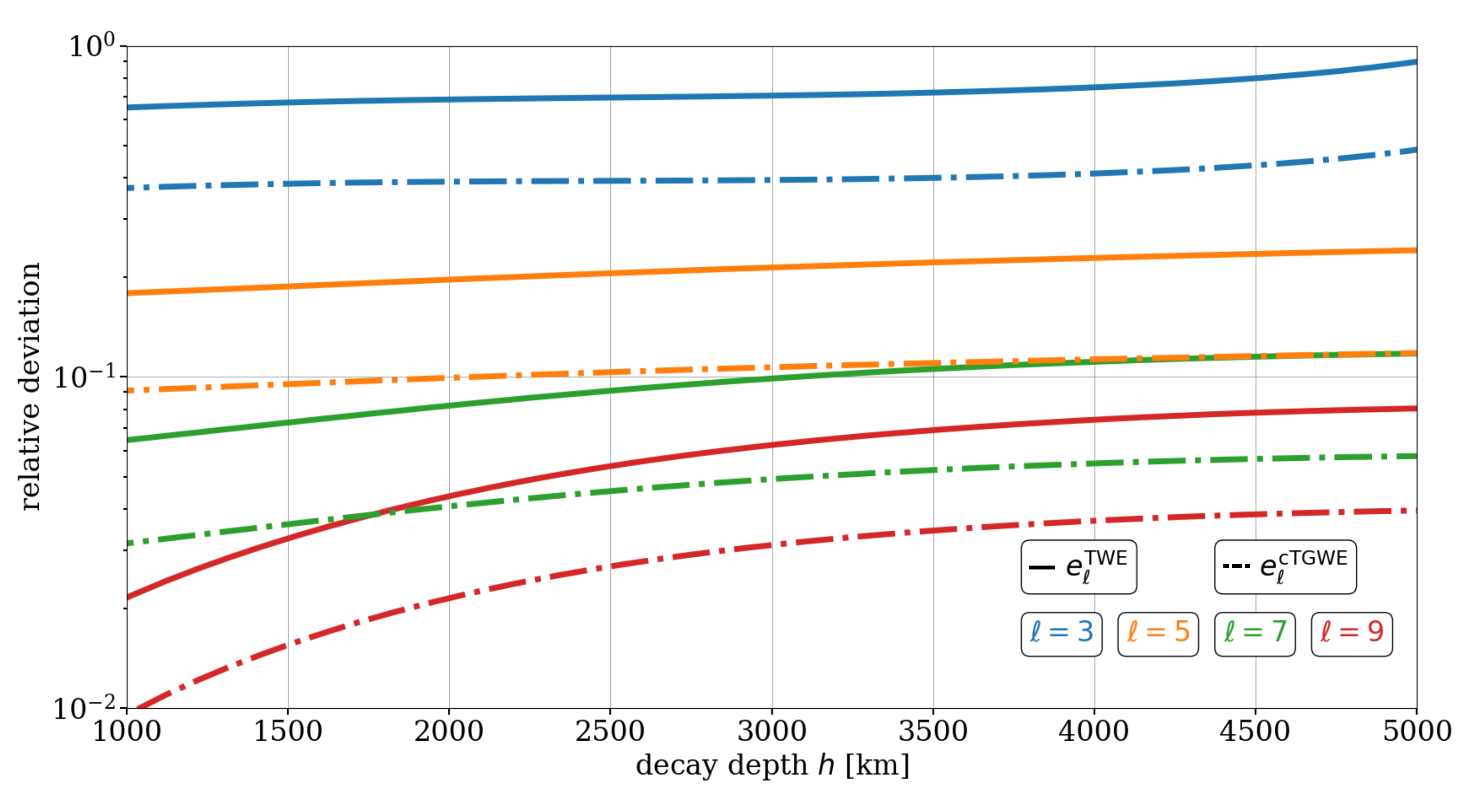}
\caption{Error of the modelled gravity moments for the TWE (solid) and constant coefficient TGWE solver (dot-dashed) relative to the reference model as function of the decay depth. Colors indicate different odd degrees.}
\label{figsolvdeg}
\end{figure}

Alternatively, we calculate the deviation of the gravity moments $J_\ell$ between different models for fixed degree $\ell$ and decay depth $h$. 
We quantify the error in $J_\ell$-error of neglecting the DSG term as a function of decay depth $h$ in terms of 
\begin{equation}
e_\ell^\mathrm{TWE}(h) = \left\vert 1-\frac{J^\mathrm{TWE}_\ell(h)}{J^\mathrm{ref}_\ell(h)} \right\vert \ .
\label{eqdTWE}
\end{equation}
The measure $e_\ell^{\mathrm{cTGWE}}$, defined in analogy to eq.~\ref{eqdTWE}, quantifies the error for the constant coefficient TGWE. Fig.~\ref{figsolvdeg} shows how both errors change for our reference model when varying $h$ between $1000\,$km and $5000\,$km (we exclude larger $h$ values because here $J_2$ changes sign
and the definition in eq.~\ref{eqdTWE} becomes useless). Except for $\ell=9$ the errors depend little on $h$ in this range. For $\ell=3$ the error $e_3^\mathrm{TWE}$ always exceeds $60$\%. For $\ell=5$ the error is still significant at about $20$\%. For $\ell=7$ and $\ell=9$, the error reaches about $10$\% for larger values of $h$. These errors are even larger than quantified by \citet{Zhang2015} or \citet{Wicht2020} and show that the DSG term with a radius dependent DSG coefficient should not be neglected. The cTGWE approach, i.e. where $\mu$ is a constant, captures roughly half of the effect of the full TGWE approach. For example, for $\ell=3$ the error $e_3^\mathrm{cTGWE}$ amounts to 40\%, whereas $e_3^\mathrm{TWE}$ exceeds 70\% at $h=3000\,$km

We can conclude that the DSG term is a first order effect and should be taken into consideration in its radius dependent form for gravity inversion problems.

\subsection{Sensitivity to surface zonal flow model}
\label{secflowprof}

We have discussed above that the different zonal wind models published over the years likely are likely to reflect seasonal variations, which may be restricted to the very shallow atmosphere and therefore do not affect the gravity signal. This motivates us to use a mean flow averaged over the available profiles from different epochs (see sec.~\ref{secaveflow} and fig.~\ref{figWD_AS}). 

\citet{Kaspi2018} relied on the HST observations obtained during Juno's perijove 9 pass \citep{Tollefson2017}, while \citet{Kong2018} based their analysis on the Cassini measurements \citep{Porco2003}. We implement both, the Perijove and the Cassini surface flow model, in our standard setup and additionally apply the amplitude cap by \citet{Kong2018} in a model we call `cap`. Tab.~\ref{tabmodres} lists the change in $h_\ell$ required when using either flow and shows that the largest influence is again found for degree $\ell=3$, where in particular the Cassini measurements require deeper sources. The Perijove profile, on the other hand yields larger offsets in the higher degrees. 
The results of \citet{Kong2018} are based in a modified version of the Cassini profile \citep{Porco2003}, i.e. capping the large spike at $20^\circ$-latitude to a maximum amplitude of $U_\mathrm{spike} = 43.2\,\mathrm{m/s}$. This flow modification is a necessary part of their model in order to match the gravity moments. Tab.~\ref{tabmodres} show that the major effect is once more a reduction of $h_3$ and a smaller increase for the higher degrees. Note once again  that the successful reproduction of Jupiter's odd-degree gravity moments by \citet{Kong2018} is due to a combination of applying this particular flow modification, the additional linear $z$-dependence, the smooth radial decay function, assuming a polytropic interior and solving the TGWE.

Generally, the impact of different surface flow profiles reaches $10$\% and is thus smaller than, for example, the effect of the equatorial step or neglecting the DSG.

\subsection{Influence of interior model}
Finally we also test the impact of using different interior structure models.
We replace the model by \citet{Guillot03} in our reference setup by the simple polytrope of index unity or the model by \citet{Nettelmann2012}.
 Details of the models are outlined in sec.~\ref{secinterior}. The chosen interior model factors in the dynamic density source (eq.~\ref{eqsourcefin}) as the background density and gravity, $\bs{\rho}$ and $\bs{g}$, and in the DSG coefficient $\mu$ (eq.~\ref{eqcoeff}). Fig.~\ref{figJinter} shows the relevant profiles of density, density gradient and $\mu$. Tab.~\ref{tabmodres} shows that the impact on the $h_\ell$ is even smaller than the impact of the surface flow model, always remaining below $10$\%. This indicates that the odd-degree gravity moments are merely insensitive to details of the Jupiter's interior structure, such as properties of the core or the heavy element abundance.

\begin{table*}
\centering
\caption{List of the models with the different treatment of the equator, different solvers, interior models and surface flows. Further given is the decay depth $h_\ell$ at which the modelled gravity moment agrees with the observations. }
\label{tabmodres}
\renewcommand{\arraystretch}{1.5}
\begin{tabular}{ccccccccc}
name & equator  & solver & interior & surface flow  & $h_3$ [km] & $h_5$ [km]& $h_7$ [km]& $h_9$ [km]    \\
\hline
{\bf reference} & full& TGWE & G03 & mean  &  $6541^{+14}_{-13}$ & $1556^{+43}_{-40}$ & $1923^{+85}_{-81}$ & $2270^{+248}_{-302}$ \\
\hline
{\bf hemispheric}& hemispheric & TGWE & G03  & mean  & $3179^{+7}_{-7}$ & $1597^{+35}_{-34}$ & $1932^{+82}_{-79}$ & $2280^{+263}_{-304}$ \\
{\bf linear z} & linear z& TGWE & G03  & mean & $5127^{+23}_{-25}$ & $1735^{+42}_{-44}$ & $2143^{+106}_{-102}$ & $2614^{+369}_{-404}$ \\
{\bf TC} & TC & TGWE & G03  & TC cut  & $2950^{+94}_{-98}$ & $2895^{+81}_{-83}$ & $2883^{+135}_{-129}$ & $3433^{+404}_{-471}$ \\
\hline
{\bf TWE} & full& TWE & G03  &  mean & $5670^{+25}_{-27}$ & $1713^{+44}_{-47}$ & $1993^{+78}_{-80}$ & $2314^{+260}_{-311}$ \\
{\bf cTGWE} & full& cTGWE & G03$^\ast$  &  mean & $6190^{+18}_{-18}$ & $1630^{+41}_{-41}$ & $1956^{+87}_{-82}$ & $2294^{+250}_{-307}$ \\
\hline
{\bf perijove} & full& TGWE & G03  & HST, perijove & $6480^{+13}_{-13}$ & $1541^{+42}_{-39}$ & $2033^{+88}_{-83}$ & $2339^{+254}_{-306}$ \\
{\bf Cassini} & full& TGWE & G03  & Cassini & $7005^{+15}_{-15}$ & $1539^{+43}_{-40}$ & $1901^{+85}_{-81}$ & $2289^{+249}_{-311}$ \\
{\bf cap} & full& TGWE & G03  &capped Cassini & $6263^{+15}_{-15}$ & $1708^{+45}_{-45}$ & $2041^{+91}_{-87}$ & $2434^{+276}_{-331}$ \\
\hline
{\bf poly} & full& TGWE & polytropic  &mean & $6213^{+23}_{-23}$ & $1637^{+52}_{-54}$ & $2043^{+107}_{-102}$ & $2465^{+306}_{-370}$ \\
{\bf N12} & full& TGWE & N12  &mean &  $6215^{+18}_{-18}$ & $1426^{+47}_{-48}$ & $1791^{+99}_{-92}$ & $2219^{+279}_{-345}$ \\
\end{tabular}
\end{table*}

\section{Discussion}
\begin{table}
\centering
\caption{Jupiter's odd-degree gravity moments from our TC model ($J_\ell$) and the observations ($J_\ell^\mathrm{obs}$) alongside the 3$\sigma$ uncertainties  \citep{Durante2020}.}
\label{tabmoments}
\renewcommand{\arraystretch}{1.5}
\begin{tabular}{rcc}
degree $\ell$ & $J^\mathrm{obs}_\ell$ [$10^{-8}$] & $J_\ell$ [$10^{-8}$] \\
\hline
3 & -4.5 $\pm$ 0.33 & -4.65\\ 
5 & -7.23 $\pm$ 0.42 & -7.59\\
7 & 12.0 $\pm$ 1.2 & 12.8 \\
9 & -11.3 $\pm$ 3.6 & -7.76 \\
11 & 1.6 $\pm$ 11.1 & 2.89 
\end{tabular}
\end{table} 

For the first time, the precise measurements of Jupiter's gravity field by the Juno mission allowed quantifying the tiny undulations caused by the zonal winds. First attempts to invert the odd-degree gravity moments suggested a rather smooth decay with depth, reaching about $2\,$m/s or even $20\,$m/s at a radius of $r=0.94\,R$ \citep{Kaspi2018, Kong2018}. This seems at odds with the magnetic observations, suggesting that the speed at this depth should not exceed the centimetre per second level \citep{Moore2019}. The gradual decay is also difficult to reconcile with the results from numerical simulations and theoretical considerations. They suggest the winds should remain geostrophic in the outer part of the atmosphere until they are abruptly quenched at some depth \citep{Christensen2020, Gastine2021}.

In addition to the too smooth radial decay, the handling of the discontinuity at the equatorial plane that can arise when asymmetric surface winds are downward continued in z-direction (parallel to the rotation axis). Such a discontinuity is unphysical, but if it existed it would contribute substantially to the modelled gravity signal.
To avoid this problem, \citet{Kong2018} therefore somewhat 
arbitrarily used an additional linear variation of the zonal with $z$ to eliminate 
the discontinuity, which is obviously at odds with a largely 
geostropic flow. 
\citet{Kaspi2018} chose an alternative 
approach and simply ignore the contribution originating from the discontinuity. 
Our analysis suggest that this was essential to successfully 
model the observed odd gravity harmonics but at the expense of physical consistency. 
More recently, \citet{Galanti2020} suggest a different radial profile that seems consistent with the magnetic observations consisting of a geostrophic outer envelope and a steep rapid decay at $r=0.972\,R$. 
They also adopt the arguable approach of ignoring the equatorial discontinuity from \citet{Kaspi2018}. The differences in the findings might be caused by various different modelling assumptions regarding how the zonal flow relates to anomalies on the gravity potential, the treatment of the equatorial step, the applied surface zonal flow model and the interior model. Here we aim to quantify the impact of each of those.

We model the odd gravity moments of Jupiter by using the thermo-gravitational wind equation (TGWE) which correctly describes the flow induced perturbations of the gravity potential. The major difference to the more commonly used thermal wind equation is the neglect of one gravity force contribution, the dynamic self-gravity (DSG). So far solving the TGWE has been possible only for polytropic interior models with index unity \citet{Zhang2015, Wicht2020}, thus we introduce a method to solve a more general form that allows to use realistic interior state models and captures the important radial variation of the DSG coefficient. 

This enables us to measure the impact of the DSG term that several authors neglect in their study \citep{Kaspi2018, Galanti2017, Galanti2020}. Our results confirm that the DSG is indeed a first order effect as already predicted by \citet{Zhang2015}, \citet{Kong2018} or \citet{Wicht2020}. The gravity harmonic $J_3$ is most severely affected and changes by up to 90\% in the models explored here. The impact then decreases with spherical harmonic order and remains below 10\% for $J_9.$ Moreover, we show that impact of the radius-dependent form of the DSG coefficient is roughly twice as large as in the polytropic  (constant DSG coefficient) form of the TGWE \citep{Wicht2020}.  

Presumably the wind structure responsible for generating the gravity anomalies is of larger scale and rather stationary, whereas all of the published jovian surface zonal wind profiles are snapshots with respect to the Jovian orbital period of 12 yrs.  We thus generate an average zonal flow profile  that represent the deep flow structure more faithfully. Our results suggest that using a specific individual profile rather than the mean significantly biases the analysis.

We also quantify the impact of different interior models and find that its influence is smaller than the equatorial treatment, the DSG term or a specific flow profile. Hence constraining the properties of the deep interior structure \citep{Guillot2018}, such as core mass or composition, on the basis of the odd gravity moments seems rather challenging as they are far more sensitive to other model assumptions, such as the equatorial treatment, considering the DSG term or assumptions about the structure of the zonal winds.

Here we suggest an alternative scenario that is motivated by
the observation that the equatorially antisymmetric flow contributions 
are much weaker outside than inside the tangent cylinder (TC) in numerical
simulations. The TC is the imaginary cylinder with a radial distance from the rotation axis that is defined by the depth at the equator where deviations from a barotropic state and/or Lorentz forces enforce a sharp drop of the zonal flow. This seems consistent with the simple fact that equatorially
antisymmetric flows are necessarily ageostrophic outside the TC, while 
there is no such conflict inside the TC. In our model we therefore
considerably damp the antisymmetric flow outside the TC, smoothing 
the transition with a tangent hyperbolic function. 
This relieves us from the problem of the 
equatorial discontinuity without having to assume an unrealistic 
$z$-dependence of the flow. We thus favour a model where the antisymmetric surface flow in an equatorial latitude band of $\pm 21^\circ$ is a shallow phenomena, not contributing to the gravity signal. This, in turn, requires strong baroclinicity in the first few hundred kilometers below the cloud level generated from latitudinal gradients in temperature or composition. Interestingly, such pronounced equatorial variations in the nadir brightness temperature and the ammonia mixing ratio have been detected by the Juno mission \citep{Bolton2017}.

We use the four significant odd moments, $J_3$ to $J_9$ to individually constrain the vertical and latitudinal structure of the zonal winds. We can match the modelled gravity moments to the observations within the 3$\sigma$ uncertainties \citep{Durante2020} when using a radial 
decay profile that models a geostrophic flow inside TC for $r>0.958 R$ but then decays 
rapidly with depth (see also tab.~\ref{tabmoments}). Outside TC the flow must be shallow. Our preferred decay profile suggests that the winds are 50\% deeper than suggested by \citet{Galanti2020} and hence the geostrophic part reaches to substantially larger pressures ($830\,$kbar rather than $300\,$kbar). At $r=0.935\,R$ the profile reaches finally $10^{-3}$ what is in agreement with the constraints from \citet{Moore2019}. This depth equals also to the lower boundary of a TC touching the surface at $\pm 21^\circ$ implying that our preferred wind structure exhibits a splitting of the gravity perturbation into an independent northern and southern hemisphere, thus the contribution from the equatorial step naturally vanishes. 

In order to explain the odd gravity moments with a physically consistent handling of the flow near the equatorial plane, we find it necessary to exclude part of the observed cloud level zonal winds from the analysis by assuming that it is restricted to shallow depth. Similarly, \citet{Kong2018} had to cap the amplitude of the strong jet centred at a latitude of $\beta = 21^\circ$ N to fit the gravity moments in their model. While it is physically reasonable that deep antisymmetric winds do not exist outside the tangent cylinder (defined with the depth at the equator where deep winds drop sharply in amplitude), there is obviously no guarantee that the antisymmetric cloud-level winds inside the tangent cylinder are fully representative of deep flow and do not also comprise a shallow component. This uncertainty is probably the severest limitation for utilising the gravity anomalies to infer the precise depth extent of the zonal flow. A better understanding of the atmospheric dynamics at different latitudes in the top few hundred kilometres below the clouds is necessary to overcome this limitation.

\section*{Acknowledgements}
We thank Eli Galanti for very constructive discussions. This work was supported by the Deutsche Forschungsgemeinschaft (DFG) in the framework of the priority program SPP 1992 'Diversity of Exoplanets'. The MagIC-code is available at an online repository (https://github.com/magic-sph/magic).
\bibliographystyle{mnras}
\bibliography{starbound}

\end{document}